\documentclass[11pt]{article}

\usepackage[margin=1in]{geometry}
\usepackage[utf8]{inputenc}
\usepackage[T1]{fontenc}
\usepackage{amsmath,amssymb,amsfonts}
\usepackage{graphicx}
\usepackage{bm}
\usepackage{xcolor}
\usepackage{setspace}
\usepackage{tikz}
\usetikzlibrary{arrows.meta,graphs,shapes.geometric,shapes.misc}
\tikzset{>={Stealth[length=2mm,width=2mm]}}

\newcommand{\bv}[1]{\bm{#1}}
\newcommand{\RR}{\mathbb{R}}
\newcommand{\tx}[1]{\textrm{#1}}
\newcommand{\N}{\mathcal{N}}
\newcommand{\PP}{ \mathbf{Pr} }
\newcommand{\amax}[1]{\underset{#1}{\arg \max}}
\newcommand{\amin}[1]{\underset{#1}{\arg \min}}
\newcommand{\dt}{\delta t}
\newcommand{\ra}{\rightarrow}
\newcommand{\EE}{\mathbb{E}}
\newcommand{\mc}[1]{\mathcal{#1}}
\newcommand{\pa}{\partial}

\newif\ifshowedits
\showeditsfalse
\newcommand{\edit}[1]{\ifshowedits\textcolor{red}{#1}\else #1\fi}

\usepackage[bookmarksnumbered,raiselinks,breaklinks,colorlinks=true,citecolor=blue,linkcolor=blue,urlcolor=blue]{hyperref}
\usepackage[backend=biber,style=phys,sorting=none, url=false, hyperref=true, eprint=false, maxcitenames=2, maxbibnames=10, abbreviate=false]{biblatex}
\title{\bf Statistical physics of language change inferred from time evolving maps}

\author{
James Burridge$^{1,\ast}$, Bert Vaux$^{2}$ \\[4pt]
\small $^{1}$School of Mathematics and Physics, University of Portsmouth\\
\small $^{2}$Department of Theoretical \& Applied Linguistics, University of Cambridge\\[6pt]
\small $^\ast$Corresponding author: \texttt{james.burridge@gmail.com}
}
\date{}

\begin{document}

\maketitle

\begin{abstract}
\noindent
Language change may be studied by tracking linguistic variant frequencies through space and time. Observed frequencies may be viewed as realisations of a statistical field. We first set out an efficient cross validated method for inferring the \textit{empirical} field --- the variant frequency distribution estimated directly from data --- in our case, high resolution surveys from 20th century USA. We then derive a model of the statistical field, assuming that individuals diffuse locally, periodically migrate, and copy one another based on a form of replicator dynamics involving both bias and linguistic accommodation. We infer model parameters by comparing changes in model and empirical fields.  Our inferences demonstrate that spatial linguistic patterns can be rapidly destroyed by migration but that accommodation can create them, protect them, and in combination with diffusion, produce partially predictable pattern dynamics. The results provide the strongest evidence yet for surface tension driven phase ordering in human dialects. We quantify the model's forecasting ability up to a time horizon of twenty five years using the Kullback-Leibler divergence between forecasts and observations.
\end{abstract}


\bigskip

\section{Introduction}

Languages typically emerge from interactions between many individual speakers \cite{lab94,san05_2}.  Because the majority of linguistic exchanges are face-to-face \cite{tru86, mil85}, language patterns evolve differently in different places and amongst different social groups. The study of the resulting spatial and social variations in language is called \textit{dialectology} \cite{cha98}. 

\textit{Statistical physics} aims to explain the behaviour of physical systems in terms of the properties of their constituent particles.   A mathematical object of central importance in statistical physics is the \textit{statistical field} \cite{kar07}, which describes a localised average of the states of the particles which make up the system. The analogy between systems of many physical particles and social systems of many people (or animals) has been widely exploited  by statistical physicists \cite{cas07}. Variations within languages can be analysed using \textit{linguistic variables} \cite{cha98}. These are linguistic features which can be realised in two or more different forms, know as \textit{variants}.  Starting from a population of speakers in a geographical area, we define the "linguistic state field" to be the vector of relative fractions of speakers near to each position who use each variant. \edit{We will restrict our attention to lexical variables. Generalising beyond this class is discussed in section \ref{sec:conc}}.

We wish to formulate state field models which explain how collective linguistic patterns emerge from interactions between speakers.  Some steps in this direction have already been taken \cite{abr03,pat09,kan10,pro17,mus19,bur20,bur19,kau21,tak20,laz23}. For example, deterministic state field models have been derived, in which speakers preferentially use the variants which they hear more often \cite{bur17,bur18,bur21}. These models predict the formation of single variant domains whose boundaries --- \textit{isoglosses} --- evolve by curvature driven dynamics similar to physical \textit{phase ordering} \cite{bra02}. In the linguistic case, however, population gradients induce forces on interfaces \cite{bur17,bur18}. The predictions of these early models have been tested in a rudimentary way by comparing their equilibria with dialect maps produced by linguists. The results in England \cite{bur17} and Italy \cite{bur18}, while visually appealing, stopped short of directly inferring parameters from data, or rigorously testing predictive power.

Parameter inference procedures for non spatial stochastic linguistic variant models are well advanced \cite{mon23,mon23,bly12}. Such models often exploit the analogy between linguistic and genetic variants \cite{bax05,bly10,bly12,wri31,fis30}. A recently developed statistical field model \cite{bur26_2} exploits the same analogy, leading to a family of coupled Wright-Fisher diffusion processes. Estimating the parameters of spatial models and testing their predictions is more challenging than in the non-spatial case. First, data is harder to obtain and process. There are non-spatial corpora \cite{dav12} that track millions of words, but high resolution spatial datasets are smaller and rarer. One reason is that simultaneously collecting spatial and linguistic information from speakers requires their active participation, usually in the form of a survey.  Modern online surveys rely on gamification methods or an element of luck in order to become popular. Particularly successful examples are the English Dialect App \cite{lee18} which collected over $50{,}000$ responses, and the Cambridge Online Survey of World Englishes (the dataset we use) which contains close to $100{,}000$ responses \cite{vau00}. Inference procedures are also harder to devise for spatial models. Some progress has been made using quasi-likelihoods \cite{bur26}, or by tuning parameters to interpolate between two snapshots of spatial variant distributions \cite{bur21}.  

While \textit{metaphorical} language models are quite common, spatial models which can be fitted to real observations are rather rare \cite{laz23,xin26,tak20,kau21}. The combination of statistical inference with statistical physics style models of individual linguistic variables is in its infancy. Although a statistical field model now exists \cite{bur26_2} with parameters that can be inferred from large scale survey data, it lacks the flexibility to capture realistic movement processes, and is limited to binary linguistic variants.  

In this paper we formulate a statistical field model which incorporates realistic long range migration and arbitrary numbers of variants --- both of which are essential in order to capture real variables and their dynamics. The model combines diffusion (localised and migratory) with variant selection based on a form of stochastic replicator dynamics \cite{hof98}. \edit{Our location-to-location migration model is
calibrated independently against government county-to-county mobility data, rather than fit to the linguistic outcomes themselves.} We estimate the model's \edit{selection parameters} in two stages: we first infer \textit{empirical} statistical fields from data, then fit parameters that most closely replicate the dynamics of these fields.
Our empirical field inference method avoids over-fitting by using Gaussian process priors \cite{ras06} and provides the most detailed reconstruction of 20th century American linguistic variables currently available. Our inferred state field model provides the strongest evidence yet for the role of phase ordering in human language via linguistic "accommodation" --- speakers adapting their variants to those of their interlocutors. Without accommodation, migration rapidly destroys linguistic variation. We characterise the ability of our model to forecast future change.

\section{Inference of empirical fields (time evolving maps)}

\label{sec:MAP}

\edit{Before introducing and fitting the statistical field model (sections \ref{sec:model} and \ref{sec:inf}) we must first estimate, directly from data, how the state field itself has evolved through time.  This step is necessary for two reasons. First, since we are inferring variant frequencies over a large land area for several decades, data is too sparse to estimate variant proportions from raw counts (Figure \ref{fig:data_counts}). We therefore need to fit a smoothed, cross-validated field which pools information across space and time. We refer to this estimate as the \textit{empirical} state field. Second, the model fitting procedure of section \ref{sec:inf} requires estimates of local trends through time, which are provided by the empirical field.}

\subsection{Details of survey data}

The Cambridge Online Survey of World Englishes (COSWE) \cite{vau00} contains $\approx 100{,}000$ responses to questions such as \textit{what is your generic term for a sweetened carbonated beverage?} Popular responses (variants) were \textit{soda}, \textit{pop} and \textit{coke}.  Most respondents reported their birth date the location where they acquired their linguistic features. The survey has been running since 2007 and remains active, with birth dates of the majority of respondents (93\%) lying in our interval of interest:  $[1950, 2000]$.

To infer historical state field evolution we use the \textit{principle of apparent time} \cite{cha98,bai91}. According to this, the linguistic state of post-adolescent speakers approximates the state of the language community to which they belonged during their adolescence.   
Comparison of real and apparent time observations support its assumptions \cite{bai91,san05,lab94,san18}. \edit{Post-adolescent lexical adoption is documented \cite{bob04}; this would make our inferred rates of change conservative. The changes we observe are substantial regardless.} Treating birth year as our time variable, the linguistic state of speakers at \textit{apparent time} $t$ (the state of all speakers born in year $t$) approximates a sample from the language community taken during the \textit{real time} interval $[t,t+a]$ where $a$ is the typical age at which a speaker's language system stabilises.  

We focus on common variants, treating rarer forms as a background  field which we do not explicitly model. \edit{In Appendix~\ref{app:background} we show that the presence of this background field does not affect the form of the selection dynamics of our model, introduced in section \ref{sec:model}.} In some cases we further simplify the data by grouping similar variants (see Figure \ref{fig:roly_sun_you}). We let $K$ be the number of variants remaining after exclusion and grouping.  After defining the variants for a  question, the response of the $i$th speaker consists of  their birth year $t_i$, their coordinates $\bv{r}_i=(r_{i1},r_{i2})$ and their variant $v_i \in \{1,\ldots,K\}$. 

\subsection{Spatial discretisation}

\begin{figure}
    \centering
    \includegraphics[width=0.7\linewidth]{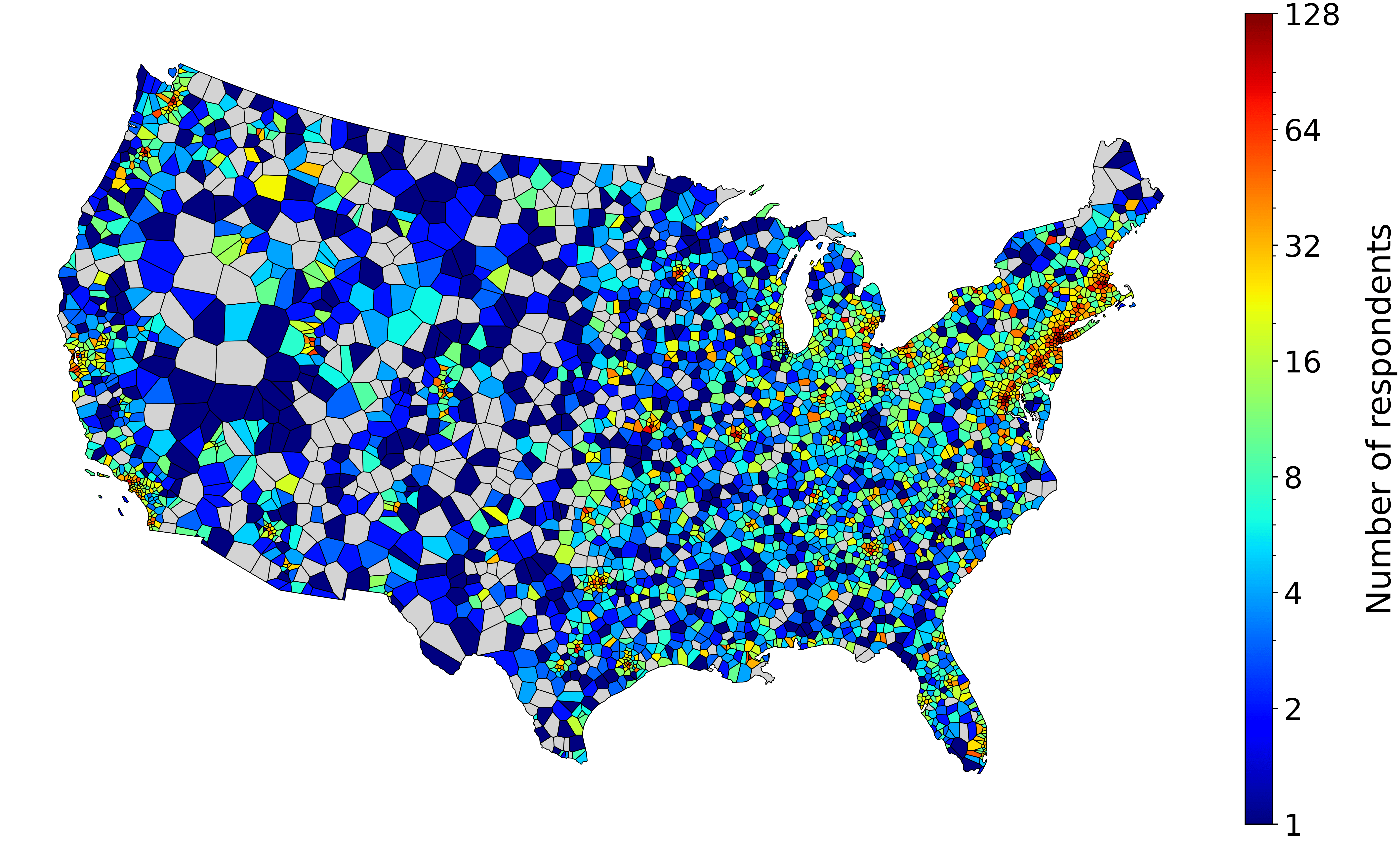}
    \caption{Respondent counts for the athletic shoes variable over the period [1950-2000]. Grey cells contain no data.   }
    \label{fig:data_counts}
\end{figure}

Our models are defined in a discrete space of cells obtained by applying the k-means algorithm \cite{has09} to all mainland USA zip code coordinates weighted by their populations as of 2020, to produce $N=4 \times 10^3$ clusters. Cells were defined by Voronoi tessellation  on cluster centroids (Figure \ref{fig:data_counts}). The mean cell population is $82 \times 10^3$ with 10th and 90th percentile cell populations $12\times 10^3$ and $207 \times 10^3$. \edit{This value of $N$ gives cells small enough so that speakers regularly interact with others outside their cells (see section \ref{sec:model}), and inferred isoglosses span multiple cells rather than being obscured by the discretisation.}

The heterogeneous population distribution is mirrored by the spatial distribution of survey responses (Figure \ref{fig:data_counts}). \edit{Approximately 10\% of cells contain no respondents for a typical variable and these cells represent approximately 2\% of the population in our sample.}

\subsection{Bayesian inference of empirical linguistic state fields}

We wish to estimate the spatial-temporal evolution of linguistic variant frequencies. For a variable with $K$ variants, we let $x_k(\bv{r}_i,t) \equiv x_{ik}(t)$ be the relative frequency with which variant $k$ is used within the cell with centroid $\bv{r}_i$ at apparent time $t$. We then define the frequency vector
$$
\bv{x}(\bv{r}_i,t) \equiv \bv{x}_i(t) = (x_{i1}(t), \ldots, x_{iK}(t))^T.
$$
The sum of relative frequencies is by definition one, so frequency vectors define a time evolving statistical field on the simplex $\Delta^{K-1}$. We call $\bv{x}(\bv{r},t)$ the \textit{state field} for the variable. To estimate the time evolution of this field given the survey data, we introduce a latent (unobservable) statistical field, 
$$
\bv{F}(\bv{r}_i,t)=\bv{F}_i(t) =(F_1(\bv{r}_i,t), \ldots, F_K(\bv{r}_i,t))^T
$$
which exists in the same space-time domain as the state field, but belongs to $\RR^K$ rather than  $\Delta^{K-1}$. To condense notation we write $F_k(\bv{r}_i,t)=F_{ik}(t)$. The state field is related to the latent field via the softmax function
\begin{equation}
\label{eqn:softmax}
x_{ik}(t) = \tx{softmax}(\bv{F}(\bv{r}_i,t))_k = \frac{e^{F_{ik}(t)}}{\sum_{j=1}^K e^{F_{ij}(t)}} .  
\end{equation}
The fact that the latent field is defined on $\RR^K$ makes it straightforward to specify a \textit{prior}, encoding assumptions about its ability to fluctuate in space and time \cite{ras06}. We use a spatial-temporal Gaussian process \cite{ras06} with four hyperparameters controlling the time-scale ($\tau$) and length-scale ($\sigma$) of fluctuations --- both via RBF kernels \cite{mur22} --- the sensitivity ($\alpha$) of length scales to population density, and the overall magnitude  of fluctuations ($\kappa$). The full definition of the prior is given in Appendix~\ref{app:MAP}.  

\begin{figure}
    \centering
    \includegraphics[width=\linewidth]{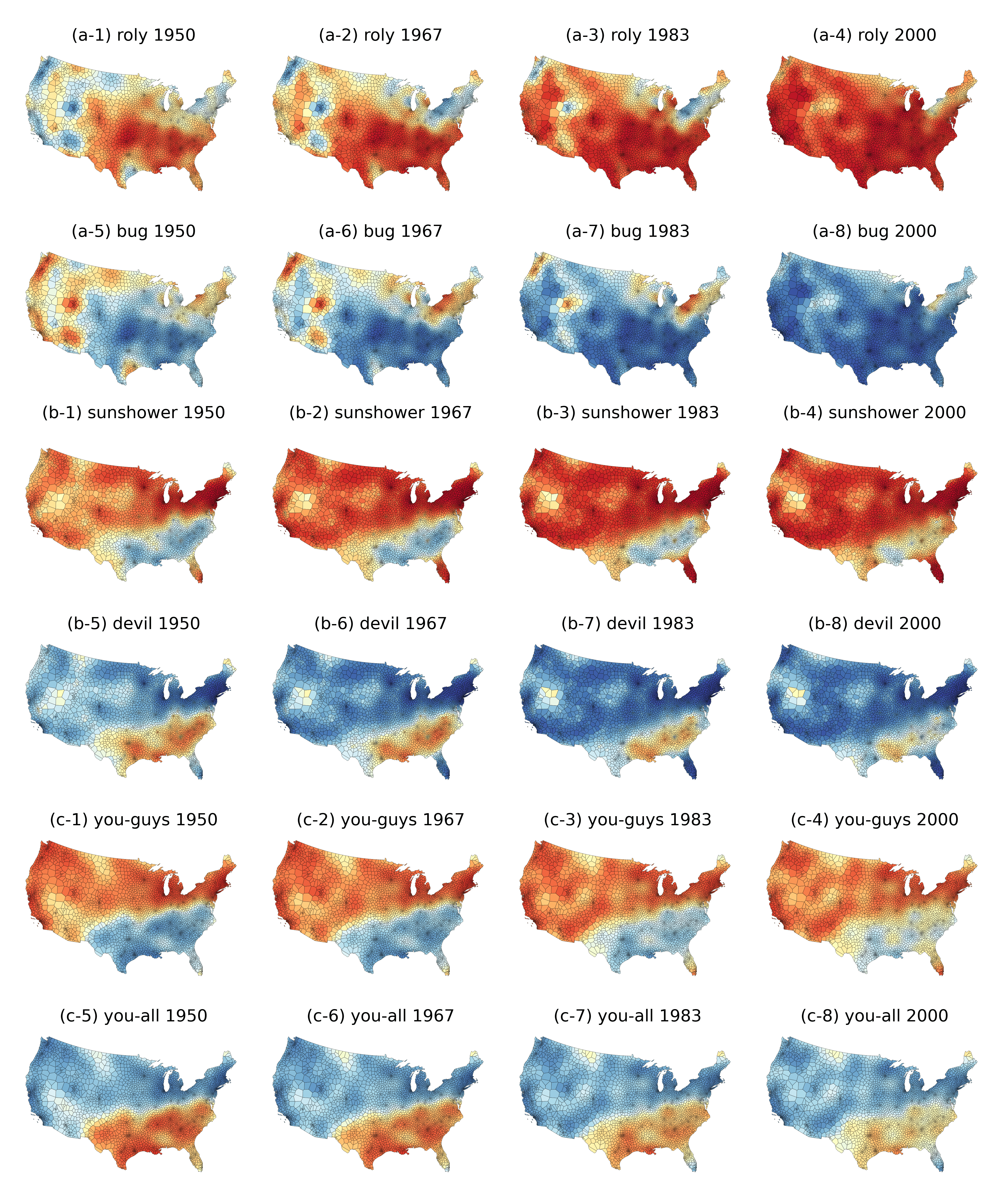}
    \caption{ Maps (a-1)--(a-8) show empirical fields for the variable "What do you call the little gray that rolls up into a ball when you touch it?". Modelled variants: \textit{roly} = "roly poly" and \textit{bug} = \{ "pill bug", "potato bug"\}. "Sow bug" (2\%) was excluded, as were respondents with no unique term for this creature (10\%). Maps (b-1)--(b-8) show fields for "What do you call the kind of rain that falls while the sun is shining?". Modelled variants:  \textit{sunshower} and \textit{devil} = "the devil is beating his wife". "Liquid sun" (1\%) was excluded, as were respondents with no variant for this question (48\%). Maps (c-1)--(c-8) show fields for  "What word(s) do you use in casual speech to address a group of two or more people?". Modelled variants \textit{you guys} = \{"you guys", "you"\} and \textit{you-all} = \{"y'all", "you all"\}. Excluded variants were "all of you" (6\%) and "you people" (2\%).   }
    \label{fig:roly_sun_you}
\end{figure}

\begin{figure}
    \centering
    \includegraphics[width=\linewidth]{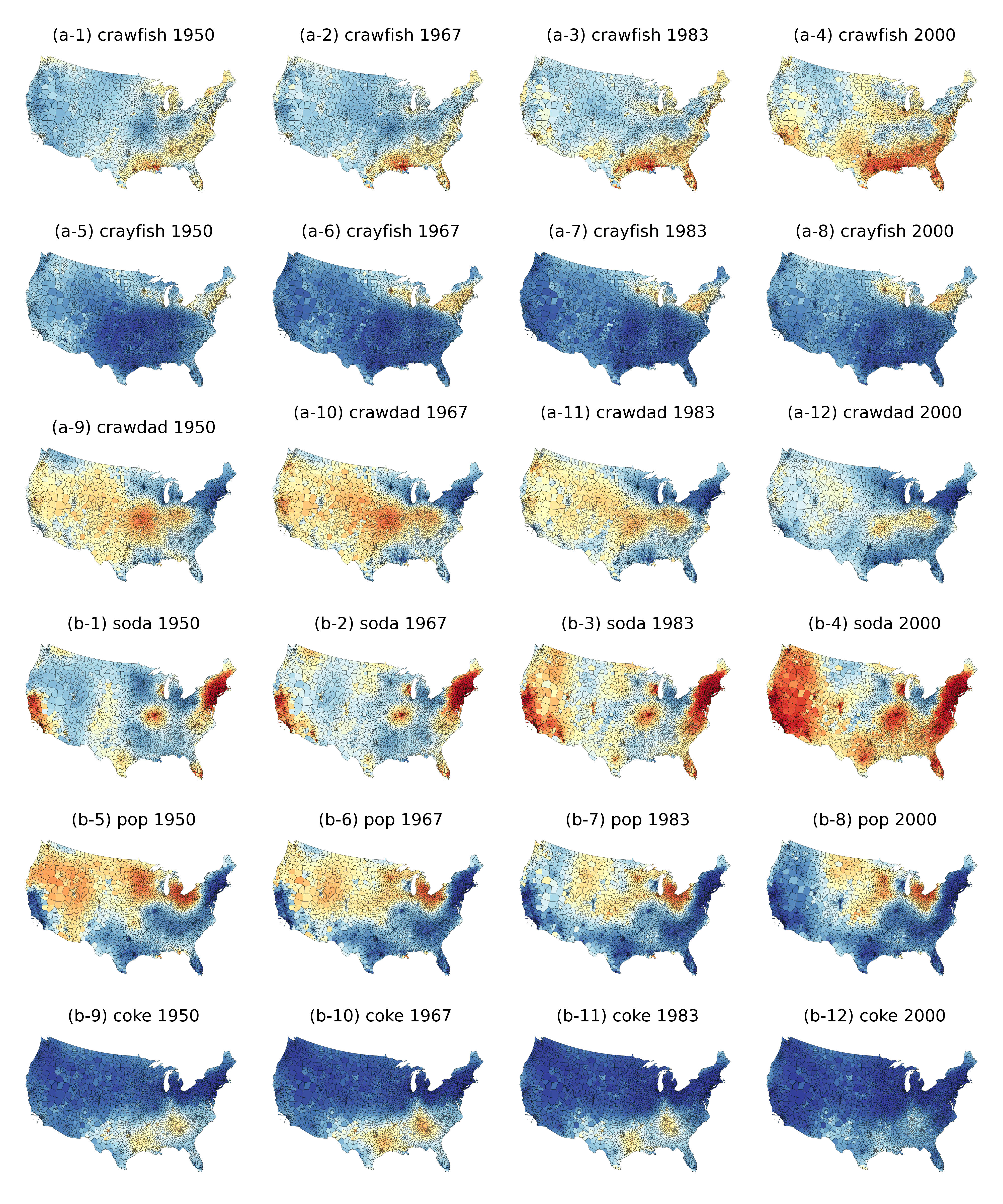}
    \caption{ Maps (a-1)--(a-12) show empirical fields for  "What do you call the kind of crustacean that looks like a tiny lobster and lives in lakes and streams?". Modelled variants: \textit{crawfish}, \textit{crayfish}, \textit{crawdad}. Excluded variants ("lobster", "crab", "prawn") have frequencies $\leq 2\%$. Maps (b-1)--(b-12) show empirical fields for "What is your generic casual or informal term for a sweetened carbonated beverage?". Modelled variants: \textit{soda}, \textit{pop}, \textit{coke}. Excluded variants were "soft drink" (7\%), "soda-pop"(2\%), "fountain drink" (1\%). Respondents with no generic term were also excluded. }
    \label{fig:craw_soda}
\end{figure}

We estimate the state field by first calculating the maximum a-posteriori (MAP) form of the latent field, then applying transformation (\ref{eqn:softmax}). We write the sequence of birth years (apparent times) in our period of interest as a vector $\bv{t}=(t_1, \ldots, t_T)$. Let $F$ be the tensor with components $F_{ijk}=F_{ik}(t_j)$, and $y_{ijk}$ be the number of respondents in cell $i$ with birth year $t_j$ using variant $k$, and let $Y$ be the tensor with components $y_{ijk}$. The probability of observing $Y$ given $F$ (the likelihood) is then
$$
p(Y|F) = \prod_{i=1}^N \prod_{j=1}^T n_{ij}! \prod_{k=1}^K \frac{\tx{softmax}(\bv{F}_i(t_j))_k^{y_{ijk}}}{y_{ijk}!}.
$$
Writing the prior density of the latent fields as $p(F)$ then the posterior density satisfies $p(F|Y) \propto p(Y|F)p(F)$ and we estimate the latent fields as
$$
\hat{F} = \amax{F}\ p(Y|F) p(F).
$$
The optimal hyperparameter vector $\bv{\psi} = (\tau, \sigma, \alpha, \kappa)$ is selected by first performing a random 80-20 train-test split of the survey data, then maximising (with respect to $\bv{\psi}$) the log probability of the test data according to field estimates obtained from the training data. Results for the six variables studied may be found in Figures  \ref{fig:roly_sun_you}, \ref{fig:craw_soda}, and \ref{fig:tennis}. We map frequencies to colours using the following scale
$$
\includegraphics[width=0.6\textwidth]{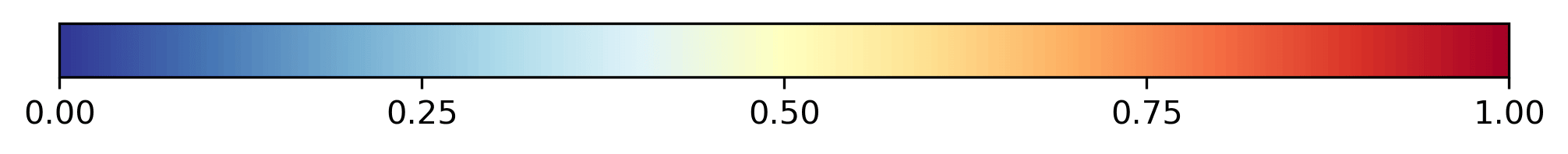}
$$
The same scale is used in all state field maps throughout the paper. \edit{Posterior uncertainty in $\hat{F}$ is higher in cells with few respondents. The Gaussian process prior mitigates this by pooling information from better-sampled neighbouring cells, so that out-of-sample predictive power is maximised.}

\section{Statistical physics of the state field}

\label{sec:model}

We now derive a model of the state field from a model of individual speakers and their interactions. 

\subsection{Diffusion}

\label{sub:diff}

Each speaker carries within them a set of linguistic variants. The variants carried by a speaker can be transported around physically if the speaker migrates from one location to another, or they can be transmitted from one speaker to another via copying. 

Let $\bv{V}_{is}(t) \in \{\bv{e}_1, \ldots, \bv{e}_K\}$ be the random unit vector indicating which variant that speaker $s$ in cell $i$ uses,  where $\bv{V}_{is}(t)=\bv{e}_k$ (the $k$th unit vector) indicates they are using the $k$th variant.  The random state field is defined
$$
\bv{X}_i(t)=\frac{1}{P_i}\sum_{s=1}^{P_i}\bv{V}_{is}(t)
$$
where $P_i$ is cell population.

\begin{figure}
    \centering
    \includegraphics[width=\linewidth]{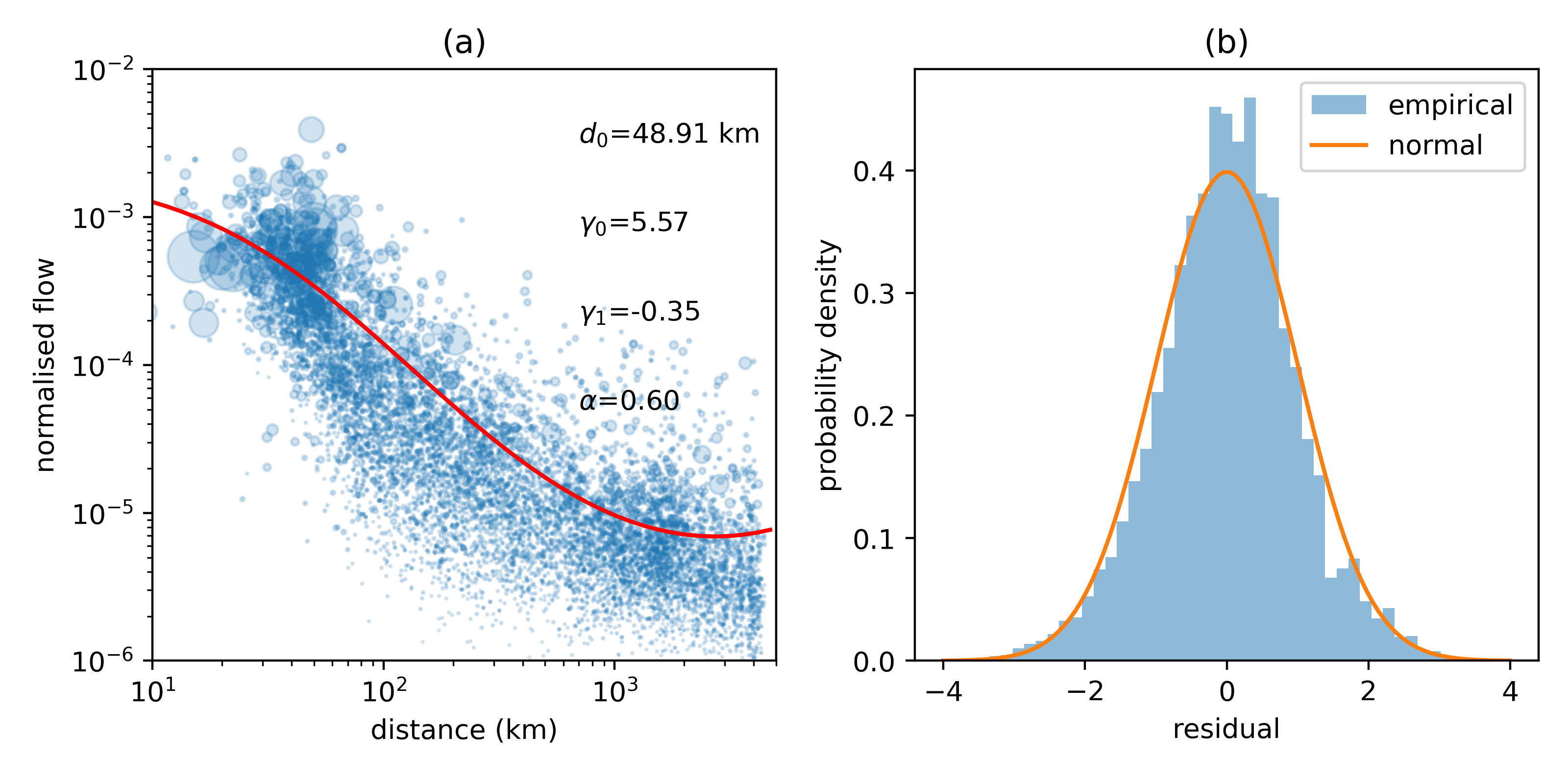}
    \caption{ (a) Normalised 2011 county-county migration flow rates. Point sizes proportional to absolute flow rates. Red curve shows inferred model, with parameter values annotated. (b) Residuals between the observed and modelled log normalised flows.  }
    \label{fig:mig}
\end{figure}

One way in which the state field can change is by speakers migrating between cells.  The distance between cells located at $\bv{r}_i$ and $\bv{r}_j$ is defined $d_{ij} = \Vert \bv{r}_i-\bv{r}_j\Vert$. In time $\dt$ we write the probability that an individual migrates from $\bv{r}_i$ to $\bv{r}_j$, where $i \neq j$, as $w_{ij} \dt$ where $w_{ij}$ is the single speaker transition rate. We assume $w_{ij}$ obeys the following gravity-type model \cite{zip46,sim12}
$$
w_{ij} = c P_i^{\alpha-1} P_j^\alpha (d_0 + d_{ij})^{-\gamma_0 - \gamma_1 \log(d_0+d_{ij})},
$$
with parameters $c>0$, $\alpha \in [0,1]$, $d_0>1$, $\gamma_0>0, \gamma_1 \in \RR$. We ignore within-cell migration since it does not change the state field. Hence we set $w_{ii}=0$. The total migration rate per speaker out of cell $i$ is $ w_i = \sum_j w_{ij}$, with transition rates between cells depending on their separations and their respective populations. The flow rate of migrants from $\bv{r}_i \ra \bv{r}_j$ is $f_{ij}=P_i w_{ij}$.
 
We estimate parameters by minimising the squared residuals (weighted by flows) between empirical and model estimates of the logarithms of \textit{normalised} flow rates $y_{ij}=f_{ij}/(P_i P_j)^\alpha$, where empirical flows are extracted from county-county migration data \cite{irs11,cen23}. Details of the fitting procedure are given in Appendix~\ref{app:migration}, and the results are displayed in Figure \ref{fig:mig}. We choose the constant $c$ so that the model matches historical migration rates. The average rate per person is given by
$$
\lambda =  \frac{1}{P} \sum_{i,j}  f_{ij}
$$
where $P$ is total population. During the period [1950,2000] (excluding years 1972–75 and 1977–80 for which the CPS 1-year mobility question was not asked) this rate remained stable \cite{cen24} with mean $\bar{\lambda} = 6.3\%$ (per person per year probability of moving county) and standard deviation $0.4\%$. We select $c=3{,}805$ so that $\lambda=\bar{\lambda}$.  

\begin{figure}
    \centering
    \includegraphics[width=\linewidth]{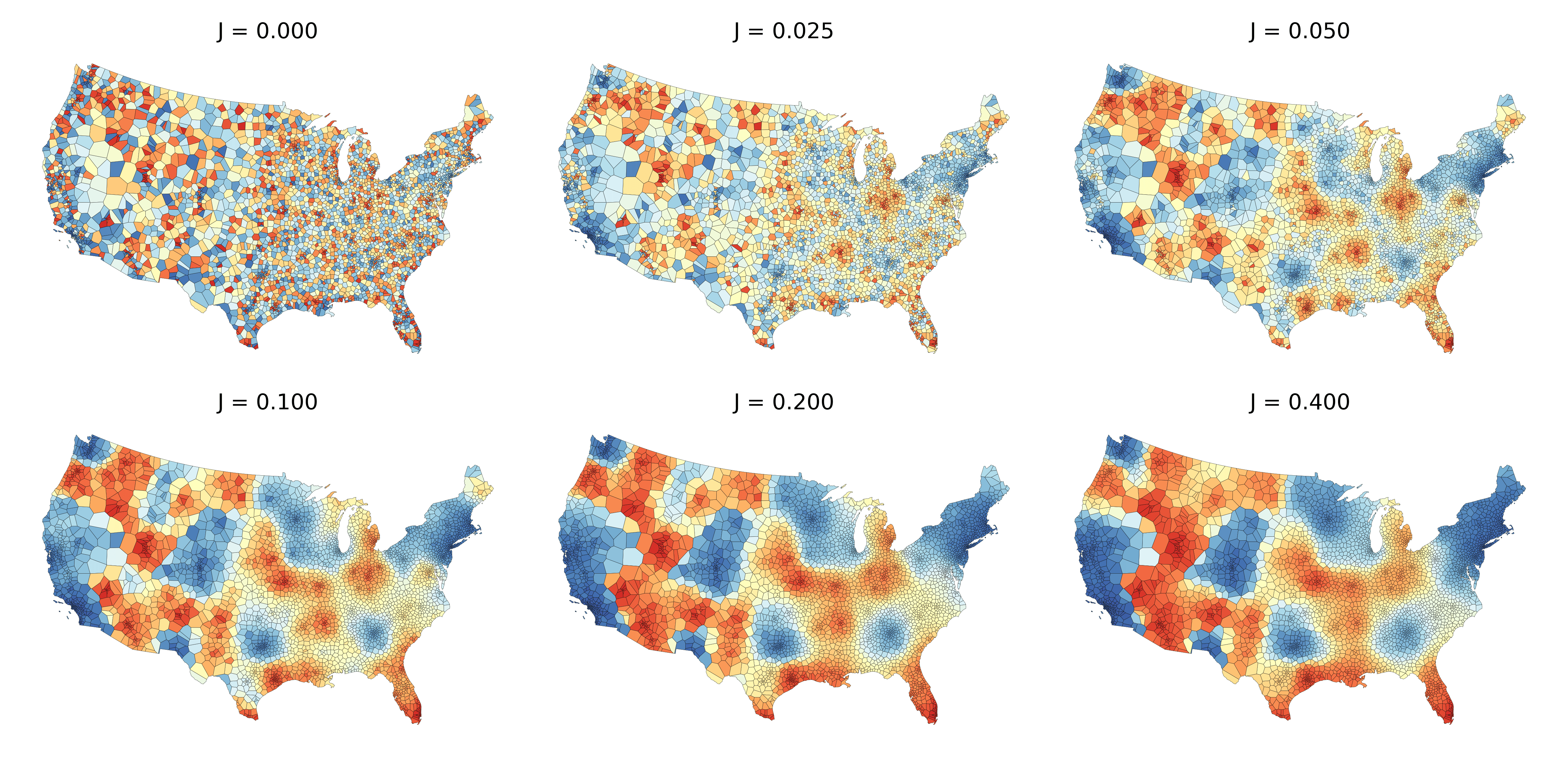}
    \caption{Result of 50 years of evolution from "primordial" state (state fields selected uniformly at random in each cell) using $\beta=0.2$ (see section \ref{sec:model}\ref{sub:sel}). }
    \label{fig:prime}
\end{figure}

To model cell-to-cell variant diffusion via copying we assume that in time $\dt$, each speaker in cell $i$ will copy the variant of a speaker selected uniformly at random from cell $j$ (possibly equal to $i$) with probability $J l_{ij} \dt$, where
$$
l_{ij} = \frac{ P_j \exp\left(-\frac{d_{ij}^2}{2R^2}\right)}{\sum_k P_k \exp\left(-\frac{d_{ik}^2}{2R^2}\right)}.
$$
According to this definition, speakers are more likely to copy variants from more densely populated cells which are near to them. The length scale $R$ is set equal to a small multiple of the typical cell radius $\bar{R}_{\tx{cell}}=25$km (we set $R=100$km).

The strength of variant diffusion due to this copying process is determined (in the limit of small cells) by the derived parameter $JR^2$ \cite{bur26_2}. Hence the choice of $R$ serves only to set the typical scale of $J$.  It may be shown (Appendix~\ref{app:diffusion-derivation}) that the expected change in the state field due to migration and copying in time $\dt$ is
\begin{equation}
\EE_t(\delta \bv{X}_i(t)) = \sum_j \left( w_{ij} + J l_{ij} - (w_i + J) \delta_{ij} \right) \bv{X}_j(t) \dt
\label{eqn:dx_diff}    
\end{equation}
where $\delta \bv{X}_i(t) = \bv{X}_i(t+\dt)-\bv{X}_i(t)$ and $\EE_t$ denotes expectation conditional on the state of the system at time $t$. From this we see that increasing the rate of local copying is equivalent to increasing physical migration between nearby cells.  

The copying rate, $J$, affects the shape and dynamics of \textit{interfaces} between geographical regions within which one variant dominates. Because it predominantly affects interfaces, it cannot  be inferred by optimising model fit at all cells simultaneously.  Figure \ref{fig:prime} shows, for different values of $J$, state field evolution from a randomised initial condition, in each case using model parameters typical of those inferred from data (this includes parameters for both migration and selection---described in section \ref{sec:model}\ref{sub:sel} below). When $J=0$  the system remains highly disordered. As we increase $J$  single-variant domains appear with smooth boundaries --- isoglosses. This behaviour is also seen in physical systems which phase-separate, and is known as phase ordering \cite{bra02}. \edit{We set $J=0.1$, producing model interfaces with widths typical of the isoglosses seen empirically in Figures \ref{fig:roly_sun_you} and \ref{fig:craw_soda}, and of the same order as the widths (20--70km) of independently measured dialect transition zones \cite{jes19}.}

\subsection{Selection}

\label{sub:sel}

\begin{figure}
    \centering
    \includegraphics[width=\linewidth]{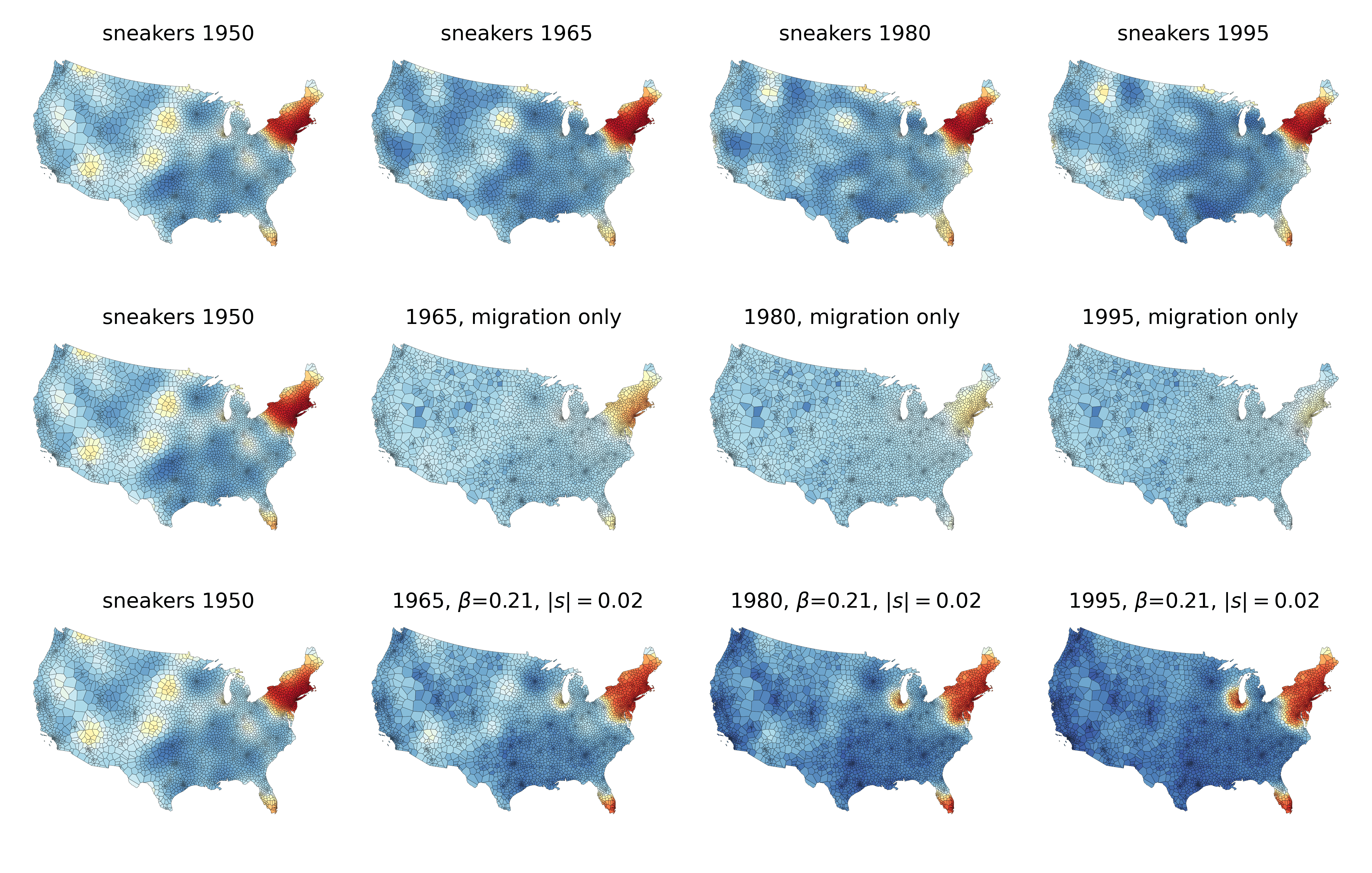}
    \caption{ Top row: empirical evolution of the \textit{sneakers} variant derived from the variable "What is your general, informal term for the rubber-soled shoes worn in gym class, for athletic activities, etc.?". Modelled variants were \textit{tennis} = "tennis shoes" and \textit{sneakers} = \{"sneakers", "sneaks"\}. Excluded variants were "gym shoes" (8\%), "running shoes" (7\%), "shoes" (5\%) with others $\leq 3\%$. Second row:  model, initialised with 1950 MAP fields, using \textit{only} diffusion. Final row:  model with same initial condition, but with bias field and accommodation factor inferred from data (see section \ref{sec:inf}).}
    \label{fig:tennis}
\end{figure}

When defining cell-to-cell variant copying we assumed that all variants were equally attractive: they have the same "fitness"  \cite{cro00,bly12}. This assumption fails to capture  important aspects of variant transmission.  First, individuals may preferentially select variants which are more common locally. Such \textit{frequency dependent selection} is known in linguistics as "accommodation" \cite{gil73}. It models the advantage of matching speech forms with you interlocutors \cite{tru04}. \edit{Individual-level accommodation has been demonstrated directly, in both naturalistic and controlled settings \cite{par11,bab10}.} Second, \textit{biases} may exist toward different variants. 
  
To illustrate the importance of accommodation, Figure \ref{fig:tennis} shows the distribution of the \textit{sneakers} variant between 1950 and 1995.  The \textit{sneakers} region has remained stable. Figure \ref{fig:tennis} also shows our model predictions if the only evolutionary forces at play were diffusive: the sneakers region would have largely disappeared. The final row shows what happens if we include accommodation and a small bias (the modelling details are given below). Here the region survives because speakers accommodate fast enough to undo the diluting effect of migrants with different variants. 
 
To model variant selection we introduce a \textit{variant fitness vector} \cite{hof98} in each cell $\bv{q}_i=(q_{i1}, \ldots, q_{iK})^T$ where $q_{ik}>0$ is the fitness of variant $k$. We assume that in time $\dt$ a speaker in site $s$ of cell $i$ who currently uses variant $j$, will transition to variant $k$ with probability 
$$
\PP(\bv{V}_{is}(t+\dt)=\bv{e}_k| \bv{V}_{is}(t)=\bv{e}_j) = \begin{cases}
    q_{ik} X_{ik}\dt &\tx{ if } j \neq k \\
    1 - (\bar{q}_i - q_{ij}X_{ij}) \dt & \tx{ if } j=k
\end{cases} 
$$
where $\bar{q}_i = \bv{q}_i^T \bv{X}_i$ is the mean fitness of all the variants in the cell. When all variants have the same fitness speakers update as if they are copying the variant of another speaker selected at random from within their cell. According to our fitness-based selection model, the change in state of an individual during time $\dt$, defined by $\delta \bv{V}_{is}(t) = \bv{V}_{is}(t+\dt) - \bv{V}_{is}(t)$, has expectation
$$
\EE_t(\delta \bv{V}_{is}(t)) = \left(\bv{q}_i \circ \bv{X}_i(t)- \bar{q}_i \bv{V}_{is}(t)\right)\dt
$$
where $\circ$ denotes the Hadamard (element-wise) product. The expected state vector increment is then
\begin{equation}
\label{eqn:rep}
\EE_t(\delta \bv{X}_i(t)) = \left(\bv{q}_i \circ \bv{X}_i(t)- \bar{q}_i \bv{X}_i(t)\right)\dt.    
\end{equation}
In the limit of large cell population, evolution becomes deterministic, and (\ref{eqn:rep}) may be recognised as the replicator equation \cite{hof98}. \edit{Replicator dynamics have previously been used to model grammar acquisition, linguistic typology, grammaticalization and lexical stress \cite{kom01,jag07,deo15,bau17}, and, in the model closest in spirit to ours, sociolinguistic convergence and divergence between discrete social groups \cite{kau20}; none treat space explicitly.} The replicator equation is invariant under the transformation 
\begin{equation}
\bv{q}'_i(\bv{x}) = \bv{q}_i(\bv{x}) + c(\bv{x}),
\label{eqn:inv}
\end{equation}
where $c(\bv{x})$ is an arbitrary scalar function. In order for the stochastic transition rates to remain positive the fitness vector must remain in $\RR_{>0}^K$, so not every scalar shift yields a well defined stochastic model. Shifts change the rate at which speakers switch variants and therefore influence the stochastic component of the dynamics, with larger fitness values leading to faster updating. We are interested only in estimating parameters which determine the expected changes, so we will allow inferred fitness values to be negative on the understanding that these must be shifted to yield a well defined stochastic model.

The fitness vector combines intrinsic bias and accommodation as follows
$$
\bv{q}_i(\bv{x}) = \bv{s}_i + \beta \bv{x}
$$
where $\bv{s}_i = (s_{i1}, \ldots, s_{iK})^T$ is the \textit{bias vector}, and  $\beta$ is the \textit{accommodation factor}. Larger $\beta$ values make it more advantageous to select variants that are more frequent in the cell, further enhancing their frequency. \edit{The accommodation and bias terms correspond to frequency-dependent and frequency-independent selection, mirroring how cultural evolution models decompose transmission bias \cite{boy85}. Similar frameworks have recently been proposed for the study of hybrid human-AI societies \cite{han26}.} \edit{In linguistic terms $s$ is a residual: whatever shapes a variant's local appeal independently of frequency, once migration and accommodation have been accounted for. It may reflect linguistic factors (ease of articulation, perceptual distinctiveness, learnability, consistency with wider grammatical or phonological patterns) or social factors such as identity, status, prestige and stigmatization \cite{lab94,lab01} --- or simply the aggregate influence of processes the model does not represent explicitly, since it is necessarily a simplified characterisation of a complex interacting system of speakers. More generally, linguistic framing can itself influence behavioural choices \cite{cap26}.}

The collection of bias vectors defines a spatially varying vector bias field which can be inferred from data. Allowing this field to vary independently between cells will lead to overfitting. To reduce the effective dimension of the field we define the bias matrix $S \in \RR^{N\times K}$ with elements $[S]_{ij} = s_{ij}$ in terms of a lower dimensional latent bias matrix $\Psi \in \RR^{Q\times K}$ with components $\psi_{ij}$. We define  the relationship between $\bv{S}$ and $\bv{\Psi}$ via a \textit{lifting matrix} $A \in \RR^{N \times Q}$ as follows
\begin{equation}
    \label{eqn:Slift}
\bv{S} = A\bv{\Psi}.
\end{equation}
The lifting matrix is constructed (see Appendix~\ref{app:lifting}) to induce spatial correlations with typical minimum length scale $\eta$. \edit{Intuitively, $\eta$ sets the smallest distance over which the bias field is allowed to vary appreciably: cells closer together than $\eta$ are pulled toward a shared bias, while more distant cells are free to differ. Larger $\eta$ therefore yields a smoother, more spatially uniform field, with the limit $\eta \to \infty$ recovering a single spatially homogeneous bias (Figure \ref{fig:pevo}). Unless stated otherwise we use $\eta = 1000$km ($Q=13$ spatial basis functions). The sensitivity of forecast accuracy to $\eta$ is explored in section \ref{sec:inf}.}

Combining diffusion and selection, we can write our model as a system of stochastic differential equations
\begin{align}
d\bv{X}_i &=  \left(\bv{q}_i \circ \bv{X}_i(t)- \bar{q}_i \bv{X}_i(t) + \sum_j \left( w_{ij} + J l_{ij} - (w_i + J) \delta_{ij} \right) \bv{X}_j(t)\right)dt + d\bv{\xi}_i(t) \\
&= \bv{b}_i\left(X;\bv{\theta}\right)dt + d \bv{\xi}_i.
\label{eqn:dX_full}
\end{align}
Here $X \in \RR^{N\times K}$ is the matrix of state field values, $\bv{b}_i$ is the drift function, $\bv{\theta}$ is the parameter vector of the model, and $d\bv{\xi}_i(t)$ is an approximately Gaussian noise increment (Appendix~\ref{app:stochastic-term}). Our focus is on inferring the parameters of the drift vector, rather than on statistics of the noise, which we set to zero when performing predictive simulations.

\section{Parameter inference, state field physics, and forecasting}

\label{sec:inf}

We infer model parameters by minimising residuals between drift increments, and increments in MAP (empirical) state fields.  We assume that real state field trajectories represent samples from our state field model for some $\bv{\theta}$. Let $\hat{\bv{X}}_i(t)$ be our MAP estimate of the state field at time $t$ and let $\hat{X}(t) \in \RR^{N \times K}$ be the matrix of state field values in all cells. The MAP field follows a stochastic process whose realised values depend on the survey data, and therefore also on the historical (but not directly observable) state field trajectories. The dependence of $\hat{X}(t)$ on the true historical process $X(t)$ over the  interval $[0,T]$ is summarised by the graph
\begin{equation*}
\tikz{
\node[ellipse, draw] (X) at (-3,0) {$\{X(t)\}_{t\in [0,T]}$};
\node[circle, draw] (Y) at (0,0) {$Y$};
\node[ellipse, draw] (hXt) at (3,0) {$\{\hat{X}(t)\}_{t\in [0,T]}$};
  \graph  { (X)->(Y), (Y)->(hXt)}; 
  }
\end{equation*}
where $Y$ is the tensor of variant counts. We assume that increments in the MAP state field may be expressed as
\begin{equation}
\delta \hat{\bv{X}}_i(t) = \bv{b}_i(\hat{X}(t);\bv{\theta}) \dt + \delta \bv{\mc{E}}_i(t),
\label{eqn:MAPinc}    
\end{equation}
where the residual increment $\delta \bv{\mc{E}}_i$ includes contributions from the stochastic component of the state field process and sampling noise in the MAP estimate, and satisfies the moment condition $\EE(\delta \bv{\mc{E}}_i(t))\approx 0$ where the expectation is taken over state field and sampling process. We do not expect noise increments to be independent. Since each MAP field noise increment is realised only once, we replace the moment condition  with an empirical analogue --- the mean residual over all cells and time increments
$$
\frac{1}{NT}\sum_{i=1}^N \sum_{j=0}^{T-1} \left(\delta \hat{\bv{X}}_i(j\dt) - \bv{b}_i(\hat{X}(j \dt);\bv{\theta}) \dt \right) \approx 0
$$
where $T$ is the number of time increments required to cover our time interval of interest. We find $\bv{\theta}$ such that this condition is satisfied as closely as possible by minimising the regularised mean squared residual
$$
\hat{\bv{\theta}} = \amin{\bv{\theta}} \left(\frac{1}{NT}\sum_{i=1}^N \sum_{t=1}^T \Vert \delta \hat{\bv{X}}_i - \bv{b}_i(\hat{X};\bv{\theta}) \dt \Vert^2 + \frac{\gamma}{QK} \sum_{j=1}^Q \sum_{k=1}^K \psi_{jk}^2 \right)
$$
where $\gamma \approx 10^{-4}$ is a small penalty which regularises the latent bias field components and imposes the soft constraint that their mean is zero. The MAP fields may be understood as being derived from the state fields via a combination of smoothing and noise (the extent of smoothing is determined by the priors). In equation (\ref{eqn:MAPinc}) we have assumed that the state and MAP fields have the same drift. This is justified provided that drift function varies approximately linearly over the time and distance scales of the smoothing.

\subsection{Migration, accommodation and interfaces}


To understand how accommodation induces and maintains interfaces like those seen in Figure \ref{fig:tennis}, we consider the continuum limit of our model along a single spatial direction parameterised by spatial coordinate $z$, assuming uniform population density and $K=2$ variants, which we will call A and B. We assume the state field is invariant along the direction orthogonal to $z$. To simplify notation we write the frequency, $X_1$, of variant A at displacement $z$ and time $t$ as $x(z,t)$.  We write the bias vector as $\bv{s}=(s,-s)$, the fraction of incoming migrants using variant A as $\bar{x}(z,t)$ and the individual migration rate as $\lambda$. We then have
\begin{equation}
\frac{\pa x}{\pa t} = D \frac{\pa^2 x}{\pa z^2} + \lambda(\bar{x}-x) + 2s x(1-x) + \beta x(1-x)(2x-1)     
\label{eqn:nld}
\end{equation}
where $D = JR^2/2$ is an effective diffusion coefficient \cite{bur26_2}. 

In the absence of bias and accommodation ($s=\beta=0$), the diffusion and migration terms will equalise variant usage throughout the system, leaving it in a spatially uniform state. In linguistics this process is referred to as "levelling" \cite{tru86}.  Now consider a spatially homogeneous system in which bias and accommodation are present. Spatial homogeneity implies that  $\bar{x}=x$, yielding dynamics 
$$
\dot{x} = 2s x(1-x) + \beta x(1-x)(2x-1).
$$

Provided that $|s|<\beta/2$ then this admits two stable steady states at $x=0$ and $x=1$ and an unstable equilibrium at $x^\ast=(\beta-2s)/(2\beta)$. The system will converge either to $x=1$ or $x=0$ depending on whether $x(0)>x^\ast$ or $x(0)<x^\ast$. 

When $s=0$, then (\ref{eqn:nld}) also admits a steady state solution consisting of (one or more) interfaces between regions where different variants dominate. Consider a single interface centred on $z=0$. Near $z=0$, provided there are no other interfaces within migration range, then the average state of incoming migrants will be $\bar{x} \approx 1/2$. Defining $x=\tfrac{1}{2}+u$ then steady state solutions of (\ref{eqn:nld}) satisfy
$$
D u'' +  a u - 2 \beta u^3=0,
$$
where $a=\beta/2-\lambda$. Provided $\beta > 2 \lambda$ this equation admits the solution 
$$
u(z) = \sqrt{a/(2\beta)} \tanh(\sqrt{a/(2D)} z),
$$ 
corresponding to the stable interface
\begin{equation}
x(z) = \frac{1}{2}\left( 1 + \sqrt{1- \frac{2\lambda}{\beta}} \tanh \left(\frac{1}{2}\sqrt{\frac{\beta-2 \lambda}{D}} z \right) \right). 
\label{eqn:inter}
\end{equation}
Introducing bias or more interfaces within migration range will cause the interfaces to move over time. In particular, if A is the majority variant amongst incoming migrants, then the A-B interface will move to expand the A domain unless opposed by a sufficiently large bias toward B speakers. 

The above analysis has implications for real dialect patterns. First, the condition $\beta>2\lambda$ quantifies when a stationary interface can exist. For the tennis-sneakers variable (Figure \ref{fig:tennis}) we inferred that $\beta\approx 0.2$, which is more than twice the empirically measured USA migration rate $\lambda \approx 6\%$. This is consistent with the preservation of the sneakers region provided that its interface with the tennis shoes region does not move inwards due to the effects of bias or migration. Second, equation (\ref{eqn:inter}) provides an analytical expression for the width of linguistic interfaces (isoglosses)
$$
w(D,\beta,\lambda) = \sqrt{\frac{D}{\beta-2\lambda}}.
$$
Using $\beta=0.2$, $\lambda=0.06$, $J=0.1$, $R=100$km we obtain $w \approx 80$km, about twice the average cell diameter of 44km. Inspection of the inferred tennis-sneakers state fields reveals that the isogloss is two to three cells wide, consistent with this width estimate, and with our estimate of $J$. 

If we generalise our continuum model to arbitrary two dimensional state fields --- by replacing $\pa_z^2$ with the Laplacian $\nabla^2$ --- we obtain a linguistic analogue of the time dependent Ginzburg-Landau equation for physical phase ordering  \cite{bra02, cas06}. The linguistic analogue of single physical phase domains are single linguistic variant domains, with interfaces driven by a combination of curvature reduction and population density gradients \cite{bur17}. The population gradient effect causes interface motion to be biased \textit{away} from population centres --- down the population density gradient --- meaning their linguistic features are to some extent protected from external influence. In the linguistic case accommodation plays the same role as local alignment processes in physical systems. The bias term acts like an external field, biasing interface motion, expanding domains in which the higher fitness variant dominates. The fact that the motion of interfaces evolving to minimise curvature is predictable in physical systems suggests that the evolution of linguistic domains should, to some extent, also be predictable.

\subsection{Spatial bias field and forecasting}

\begin{figure}
    \centering
    \includegraphics[width=\linewidth]{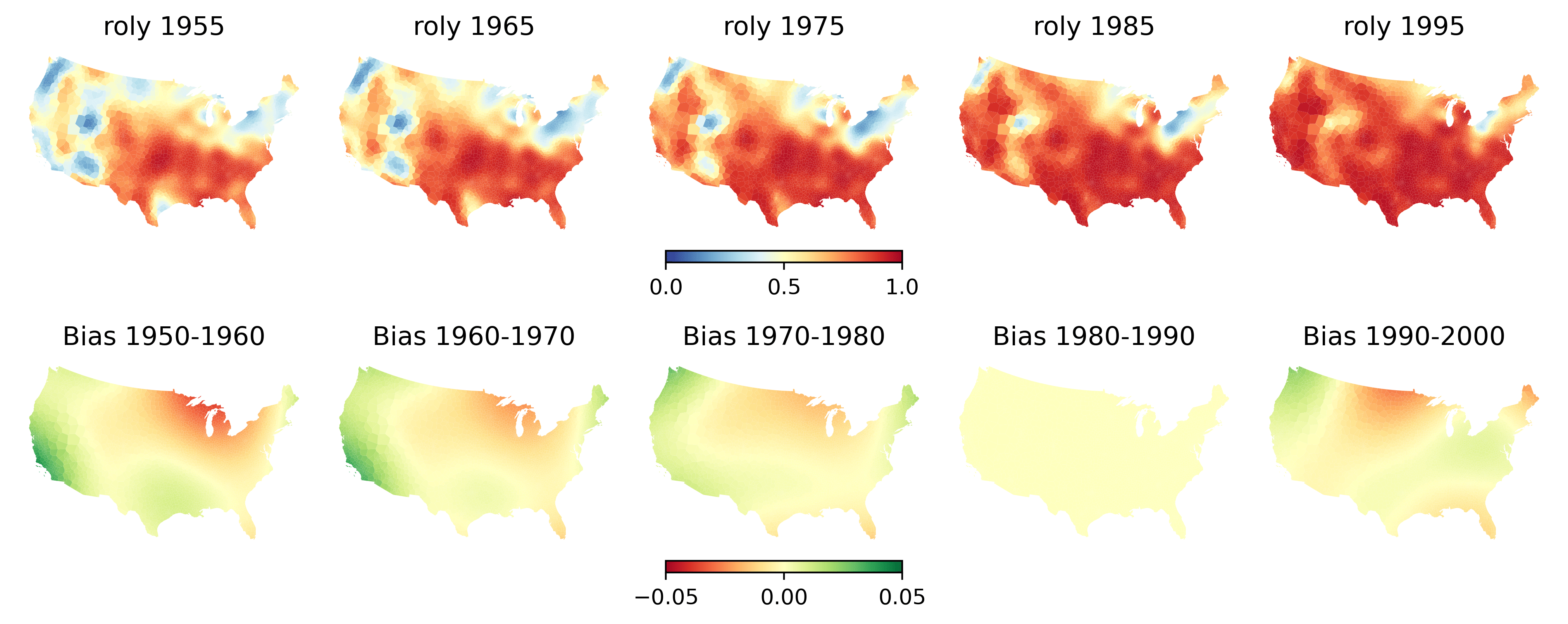}
    \caption{ Top: empirical fields for the \textit{roly} variant. Bottom: bias fields ($\eta=1000$km) inferred from empirical fields within each decade. Accommodation factors inferred for each decade are $\hat{\beta} = 0.19, 0.20, 0.19,0.15,0.10$.  }
    \label{fig:roly_bias}
\end{figure}

\begin{figure}
    \centering
    \includegraphics[width=\linewidth]{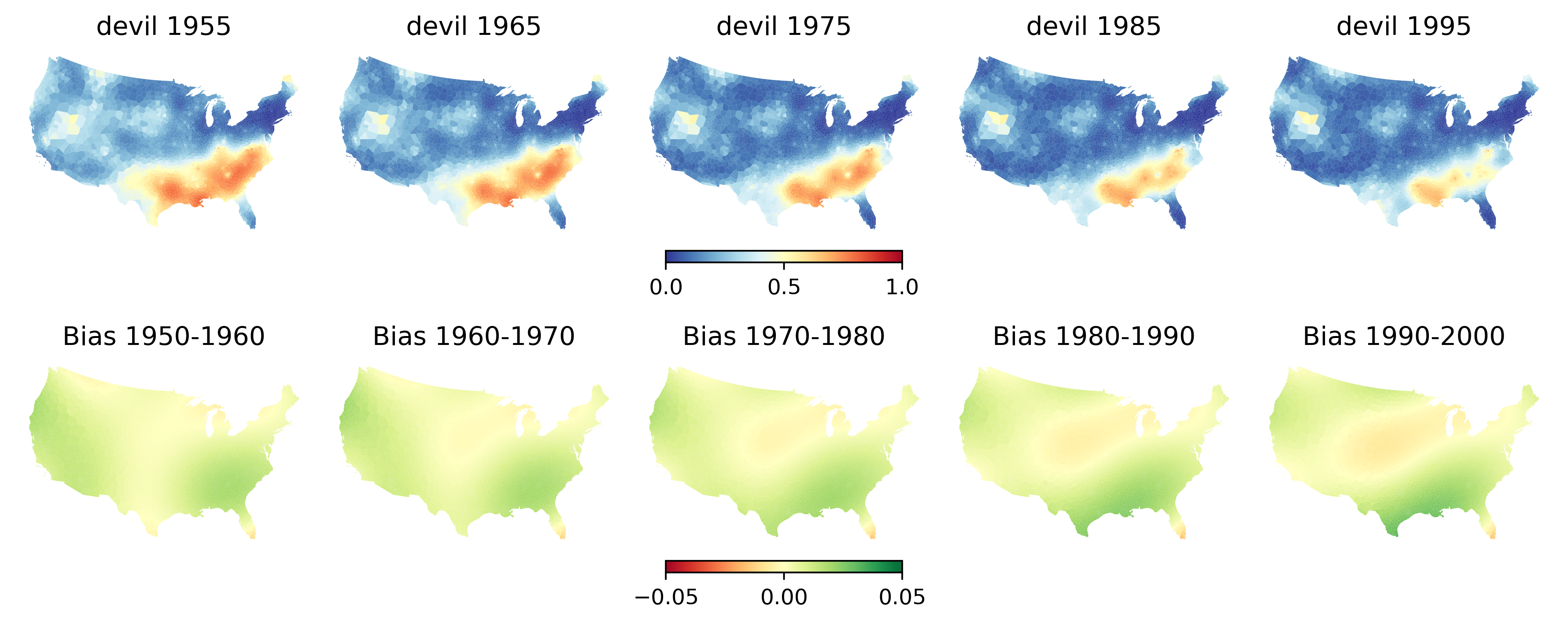}
    \caption{ Top: empirical fields for the \textit{devil} variant. Bottom: bias fields ($\eta=1000$km)  inferred from empirical fields within each decade. Accommodation factors inferred for each decade are $\hat{\beta} = 0.18, 0.17, 0.16,0.13,0.09$.  }
    \label{fig:devil_bias}
\end{figure}

We now consider the model with spatially varying bias field, fitted to variables whose distributions have changed over the study period.  Figure \ref{fig:roly_bias} shows the empirical field for "roly-poly" used to describe a woodlouse. Also shown is the bias field (length scale $\eta=1000$km) inferred for each decade in [1950,2000]. Roly-poly has expanded from the South and southern Midwest into the  North and West, eventually becoming dominant throughout the USA. From 1950-1980 the inferred accommodation factor was $\hat{\beta}\approx 0.2$ (matching tennis-sneakers), above the threshold required to maintain spatial variation when the average incoming state is an equal balance of variants. In fact the \textit{roly} variant was in the majority in 1950 (see Figure \ref{fig:pevo}), meaning that in most regions the majority of incoming migrants would have been \textit{roly} users. We might expect this to have shifted interfaces in favour of \textit{roly}, but in Los Angeles (and other cities) a combination of population gradients and geographical distance from other population centres provides a protective effect. To explain the colonisation of Los Angeles by "roly poly" using our model therefore requires a strong positive bias ($\hat{s} \approx 0.04$) in favour of the \textit{roly} variant in the Los Angeles region, sufficient to "flip" it away from the \textit{bug} variant. As \textit{roly} becomes increasingly dominant in the country as a whole, the flow of \textit{roly} speakers into \textit{bug} regions will increase, pushing the country toward the pure \textit{roly} state. 

An alternative explanation for the change in Los Angeles is that the population of Los Angeles County grew from 2.7 million in 1940 to 7 million in 1970, requiring unusually high rates of inward migration \cite{for96}. This would have resulted in a bias toward the national majority variant. In order to resolve the relative importance of migration and bias, we would need a migration model capable of reproducing the historical population movements more accurately.      

A second example of the interaction between bias, accommodation, and migration is the shrinkage of the \textit{devil} variant of the variable "What do you call the kind of rain that falls while the sun is shining?" (Figure \ref{fig:devil_bias}). Here we infer a small positive bias in its favour throughout the period of interest, and gradually declining accommodation factor. Spatial variations in bias are smaller than in the previous example. The fact that the devil domain shrinks despite positive bias in its favour may be explained by a combination of incoming speakers whose average state is closer to the \textit{sunshower} variant, and to a decline in the protective effect of accommodation. 

\begin{figure}
    \centering
    \includegraphics[width=\linewidth]{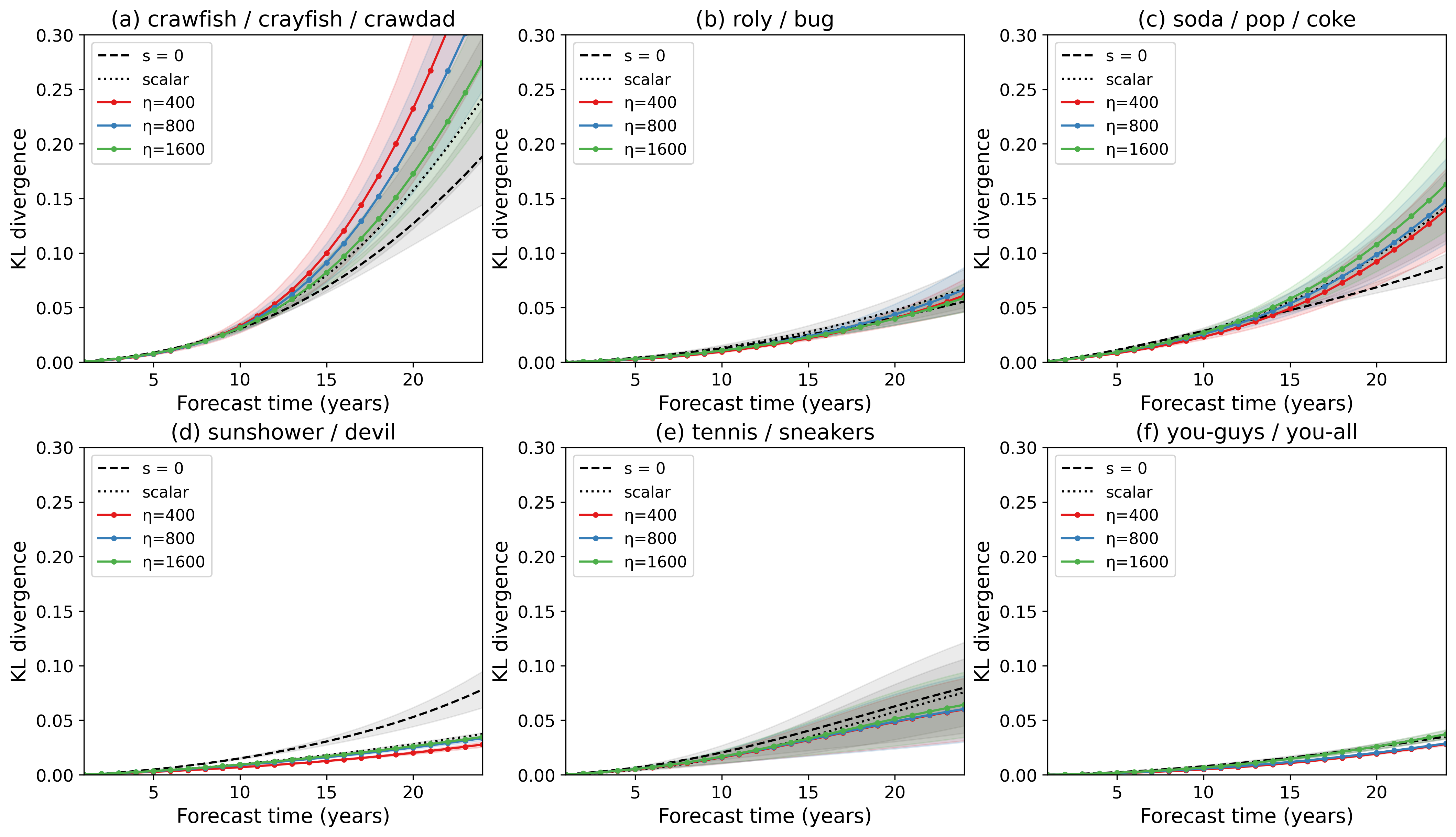}
    \caption{ KL divergence between model forecasts  and empirical fields. Bands show $\pm 1$ standard deviation in the KL divergence values over all training windows.  }
    \label{fig:KLD}
\end{figure}

\begin{figure}
    \centering
    \includegraphics[width=\linewidth]{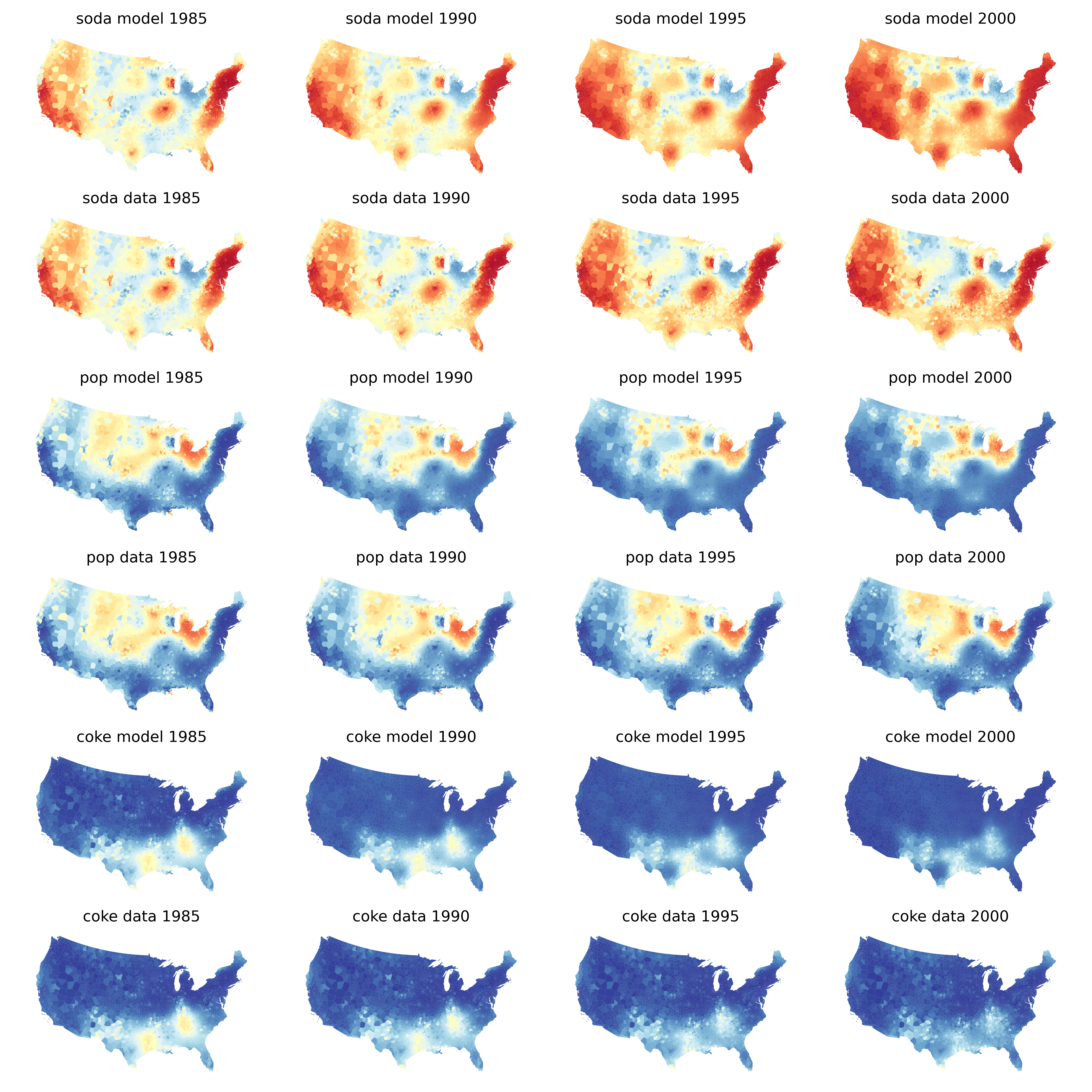}
    \caption{Forward predictions of the \textit{soda}, \textit{pop} and \textit{coke} state fields using model parameters fitted to MAP fields in the training interval [1975,1985], with $\eta=2000$km. Inferred accommodation factor $\hat{\beta}=0.20$. }
    \label{fig:spc_pred}
\end{figure}

To assess the model's ability to make predictions we infer model parameters over a decade long "training period",  initialise the model with the MAP field at the end of the period, and evolve it forward in time by solving the system of differential equations (\ref{eqn:dX_full}) with $d \bv{\xi}_i=\bv{0}$. In each cell we compute the Kullback-Leibler (KL) divergence \cite{kul51} between the predicted and MAP state fields at future time $t$
$$
D_{\tx{KL}}(\hat{\bv{x}}_i^{\tx{MAP}}(t) \Vert \hat{\bv{x}}^{\tx{pred}}_i(t)) = \sum_{k=1}^K \hat{x}^{\tx{MAP}}_{ik}(t) \log \frac{\hat{x}^{\tx{MAP}}_{ik}(t)}{\hat{x}^{\tx{pred}}_{ik}(t)}.
$$
The predictive performance of the model at time $t$ is measured by the average KL divergence over all cells at that time. Repeating the process for a series of overlapping training decades and bias field length scales allows us to estimate the mean and standard deviation of this performance measure over different prediction horizons. We also calculate the performance of the constant bias and zero bias models using the same method.  Figure \ref{fig:KLD} shows the relationship between KL divergence and forecast horizon for each of these models. In panels (b) and (f) we see minimal difference in the forecasting ability of different models; in both cases the inferred biases are close to zero (see also Figure \ref{fig:pevo}). In panels (a) and (c) all models perform similarly well initially (ten to fifteen years into the future), after which the more complex models perform worse that the zero bias model.   For reference, Figure \ref{fig:spc_pred} illustrates  5, 10 and 15 year forward predictions from the model for the \textit{soda}, \textit{pop} and \textit{coke} variants, with $\eta=2000$km, with corresponding KL divergences of 0.007 (5 year), 0.02 (10 year) and 0.04 (15 year). In panels (d) and (e) of Figure \ref{fig:KLD} the situation is reversed --- the presence of a bias field improves forecasts from five years onward. Taken together these observations suggest that when forecasting a decade into the future,  bias fields inferred from the recent past improve (or do not worsen) future predictions. However \textit{spatial variations} in bias inferred from the past do not, in general, improve future predictions.  \edit{Figure \ref{fig:KLD} varies $\eta \in \{400, 800, 1600\}$km alongside the  zero-bias and spatially constant ($\eta \to \infty$) models; forecast accuracy is not strongly sensitive to this choice, and the values used elsewhere ($\eta = 1000$km, $2000$km) lie within or just above this tested range.}

\subsection{Time evolution of parameters for all variables}

\begin{figure}
    \centering
    \includegraphics[width=\linewidth]{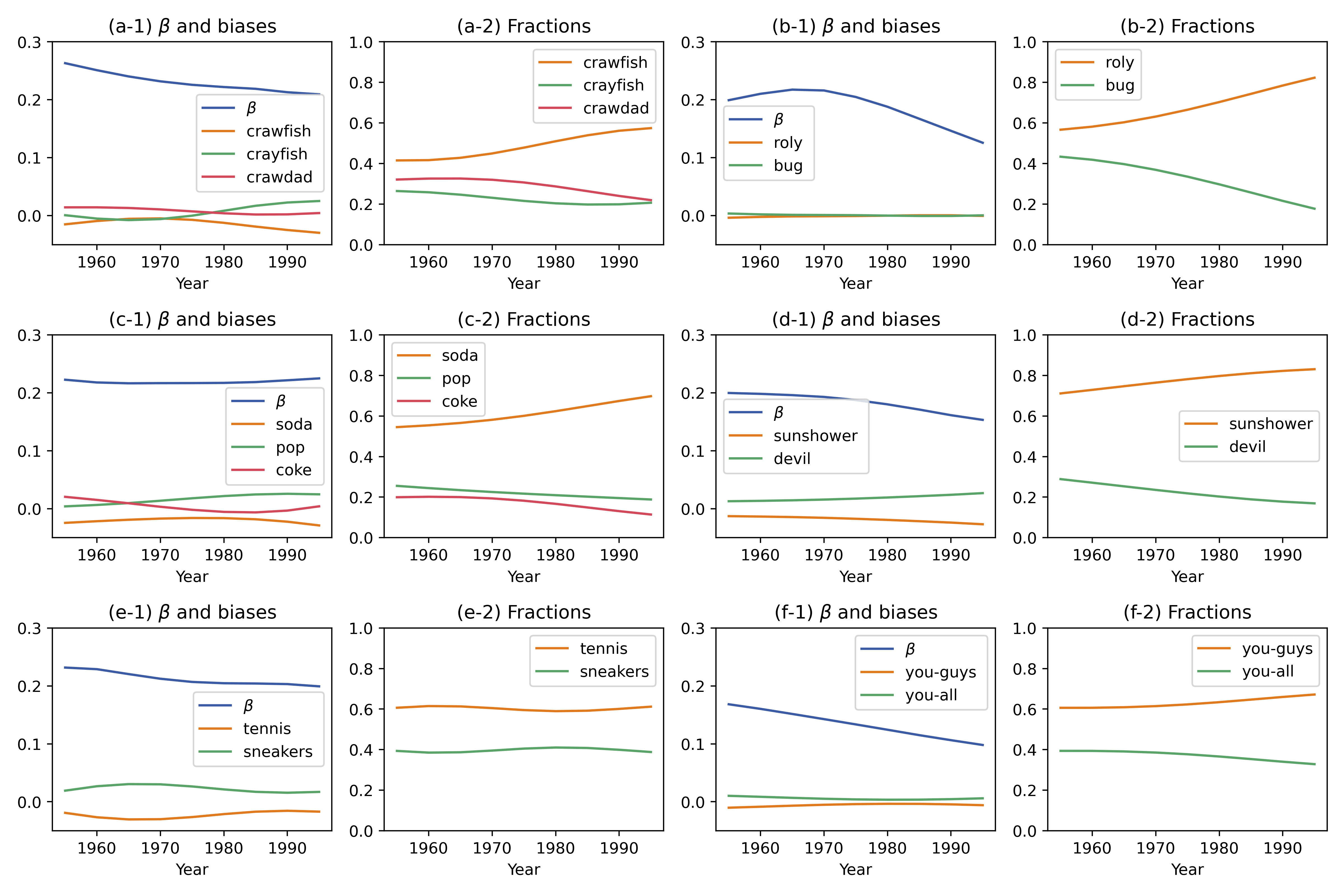}
    \caption{Panels (a-1) to (f-1) show accommodation factors and inferred spatially homogenous biases ($\eta=\infty$) for every variant of each variable. 
    Panels (a-2) to (f-2) show fractions of the population using each variant over the same time period. }
    \label{fig:pevo}
\end{figure}

Figure \ref{fig:pevo} shows accommodation factors and spatially constant biases fitted in decade-long sliding windows, together with national variant proportions. The accommodation factor is remarkably consistent across variants and through time, and comfortably large enough to prevent migration from levelling spatial variation. This provides evidence for accommodation's role in linguistic pattern dynamics, which has previously been difficult to obtain from the low-resolution spatial snapshots typically available \cite{bur17,bur21,bur26_2,tru04,bly12,cha98}.

For all but one variable we  see a small decline in accommodation over the study period, consistent with a reduction in the distinctiveness of local linguistic variations. A possible explanation for this is that interactions (as opposed to physical migration) have become less localised over time, perhaps due to greater audio-visual media influence. Since this effect is not explicitly contained in our model, it is possible that the inferred accommodation factor is shrinking to compensate.  

In four out of the six variables plotted in Figure \ref{fig:pevo} (panels (b), (c), (d) and (f)) the inferred bias for the variant which has \textit{increased} in proportion over the study period is zero or negative. This counter intuitive result may be explained by considering the combined effect of migration and accommodation. In each of these examples the growing variant --- which we'll call A --- begins in the majority. Consequently, in the majority of locations the fraction of A speakers will increase due to accommodation. Further, the majority of migrants will also be A-speakers, exerting an effective bias in favour of this variant. A negative bias inferred for an expanding variant therefore implies that it is growing somewhat slower than would be expected given migration and accommodation alone. Of course this picture is simplified: the geography of real populations can protect minority regions \cite{bur17,bur21}, and in regions where variants signify local identity, there will be a positive bias in their favour \cite{lab63}.

\section{Linguistic perspective}

\label{sec:linguistic}

We now consider the implications of our results for linguistics. The inferred state fields reveal several patterns of significant interest from a dialectological perspective. The roly poly / pill bug boundary (Figure \ref{fig:roly_sun_you}a) largely tracks the traditional boundary between the South and North Midland dialect regions established by Kurath 1949 \cite{kur49}, lending confidence to the state field inference method (section \ref{sec:MAP}): we recover a well-documented dialect boundary from survey data alone, without any prior knowledge of traditional dialect regions. The sunshower / devil beating his wife maps (Figure \ref{fig:roly_sun_you}b) are particularly striking, since the devil-based expressions, which are a regionally and socially marked form associated with Southern and African American vernacular traditions, show marked contraction over the study period despite a small inferred positive bias in their favour, suggesting that accommodation and migration are jointly overwhelming what appears to be a socially embedded prestige or identity effect. This is consistent with Labov’s distinction between markers (variables subject to social evaluation) and indicators (variables below social awareness): identity-linked forms may retain positive bias yet still recede under demographic pressure \cite{lab94,lab01,lab10}. The soda / pop / coke variable (Figure \ref{fig:craw_soda}b) illustrates a three-way competitive system in which regional boundaries have remained remarkably stable across the study period, with coke confined to the South and pop dominant across the Upper Midwest and Pacific Northwest, a pattern whose persistence through half a century of high migration is now quantitatively explained by the model (sections \ref{sec:model} and \ref{sec:inf}) via the accommodation mechanism. The sneakers / tennis shoes isogloss (Figure \ref{fig:tennis}), running roughly from Washington DC to Cleveland, is one of the sharpest lexical boundaries in American English and has long attracted attention precisely because it resists the levelling one would expect given the scale of post-war population movement in the Northeast corridor; the model’s demonstration that accommodation alone is sufficient to account for the maintenance of this boundary, without requiring any special social salience or prestige effect --- \edit{which would be modelled as a strong localised bias} --- is a significant result. 

There are a number of areas where our modelling results are relevant to ongoing debates in linguistics.  There has been a long-running debate in sociolinguistics, going back to Trudgill’s work on dialect contact and new-dialect formation \cite{tru86,tru04}, about whether accommodation suffices to explain the persistence of dialect boundaries, or whether additional factors such as social identity, prestige, and stigma are necessary \cite{bax09}. Our model provides the first quantitative demonstration, fitted to real spatial data over decades, that accommodation alone (at empirically inferred rates of $\beta \approx 0.2$) is sufficient to maintain sharp isoglosses against realistic levels of migration. This doesn’t rule out social factors, but it establishes a precise quantitative baseline.  There has been considerable debate, particularly in British sociolinguistics \cite{tru86,ker03,bri02}, about whether increased mobility necessarily leads to dialect levelling. Our model gives a rigorous condition within the unbiased two-variant interface approximation: a stationary interface exists provided the migration rate is less than half the accommodation factor ($\lambda<\beta/2$). This converts what has been a qualitative debate into a testable quantitative condition, and empirically, for 20th century USA, the condition for maintaining variation is comfortably met.  Our paper uses the apparent-time construct \cite{bai91} to infer historical evolution. \edit{Independent support for this comes from the recovery of the independently-documented Kurath dialect boundary, which subsequently changes (see above).}

\section{Conclusions}

\label{sec:conc}

We have made two methodological contributions toward the development of a statistical field theory for language. The goal of this discipline is to explain how coarse grained patterns in language use emerge from interactions between individual speakers. In particular, we want to understand the extent to which the evolution of the state field is predictable, and what kinds of collective behaviour can emerge. To address these questions quantitatively requires modelling, and models must answer to data. Our view is that statistical physics is the ideal tool to develop such models, but that this must be combined with practical methods of inference.

Our first contribution was an efficient method for inferring the (apparent time) historical evolution of state fields. We made use of Gaussian process priors and cross validation in order that our inferred fields provided a maximally plausible model of the true fields. To our knowledge, the resulting maps provide the most detailed examples of the evolution of American linguistic features over the second half of the 20th century. Our second contribution, closely related to the first, was to define a new multi-variant statistical field model which was capable of explaining the observed fields in terms of realistic individual diffusion and selection processes. These processes included migration, local diffusion, linguistic accommodation and spatially varying "biases". We inferred the parameters of our model, excluding stochastic terms, by comparing changes in our historical fields with changes predicted by the model. Our inferences showed that accommodation and migration are powerful creative and destructive forces in linguistic pattern formation, and the pattern evolution is partially predictable.   
\edit{The six variables studied here are lexical, treated as independent, discrete choices among $K$ variants. Extending to grammatical or phonological variables \cite{kom01,jag07,deo15,bau17} may require coupling between variables, generalising the order parameter itself, for example to describe position in acoustic space, and accounting for the more complex acquisition dynamics through which such variables are learned. Variables where identity maintenance drives divergence rather than convergence may require an additional field label representing group identity, with a fitness term rewarding divergence from other groups, as in \cite{kau20}.}

\section*{Data Accessibility}
The code and state-field data generated by this study are available from the public GitHub repository \url{https://github.com/james-burridge/COSWE_stat_phys}

\section*{Author Contributions}
Burridge developed the mathematical and statistical methodology. Both authors contributed to interpretation and analysis of results.

\section*{Funding}
The authors are grateful for the Royal Society APEX award \texttt{APX\textbackslash R1\textbackslash 241139} which supported this research

\appendix

\section{Details of empirical state field inference}
\label{app:MAP} 

\subsection{State field prior}

The relation between latent and state field is given by
\begin{equation}
\label{eqn:app-softmax}
x_{ik}(t) = \tx{softmax}(\bv{F}(\bv{r}_i,t))_k = \frac{e^{F_{ik}(t)}}{\sum_{j=1}^K e^{F_{ij}(t)}}.  
\end{equation}
Our prior on the latent field is a spatial-temporal Gaussian Process which is discrete in space (defined on the cells of our model) and continuous in time. A Gaussian process is a stochastic process such that every finite collection of field values --- in our case specified by a set of location-variant-time coordinates --- has a multivariate normal distribution \cite{ras06}. We define our prior by specifying the means and covariances of all such finite collections. Our first assumption is that the expected latent field under the prior is zero, corresponding to a state field with equal frequencies for all variants at all locations and times. Our second assumption is that the covariance of two latent field values for the same variant at the same location at different times $t_1$ and $t_2$ is given by the radial basis function covariance kernel \cite{mur22} with time scale $\tau$
$$
K_t(t_1,t_2) = \kappa^2 \exp\left( -\frac{(t_1-t_2)^2}{\tau^2} \right).
$$  
Here the parameter $\tau$ determines the time scale over which field values are correlated with one another in the prior. Larger values of $\tau$ lead to posterior fields which vary more smoothly --- fluctuate less rapidly --- in time.  The parameter $\kappa$ determines the magnitude of latent field fluctuations within the prior. Small $\kappa$ values lead to posterior fields which are closer to zero, and therefore closer to equal-frequency state fields. In order to introduce prior assumptions about spatial fluctuations we introduce a spatial kernel $K_s(\bv{r}_1,\bv{r}_2)$ with $K_s(\bv{r},\bv{r})=1$, and define the covariance between the latent field values for the variants $k_1$ and $k_2$ at two space space-time points $(\bv{r}_1,t_1)$ and $(\bv{r}_2,t_2)$ as
$$
\tx{Cov}(F_{k_1}(\bv{r}_1,t_1),F_{k_2}(\bv{r}_2,t_2)) = K_s(\bv{r}_1,\bv{r}_2) K_t(t_1,t_2) \delta_{k_1,k_2}.
$$
Here the Kronecker delta $\delta_{k_1,k_2}$ imposes the assumption that latent fields for different variants are uncorrelated in the prior. Posterior dependence between latent variant fields is derived from the data. Transformation (\ref{eqn:app-softmax}) also induces dependencies between state fields, due to the constraint that they belong to the $K-1$ simplex. To allow for the possibility that spatial correlations depend on population density, $\rho(\bv{r})$, we introduce the standardised log density
$$
z_i = \frac{\log \rho(\bv{r}_i) - \langle \log \rho \rangle}{\sqrt{\langle (\log \rho)^2 \rangle - \langle \log \rho \rangle^2} }
$$
where $\langle \bullet \rangle$ denotes the average of $\bullet$ over all cells. Using this measure we define the adjusted length scale for cell $i$
$$
l_i = \sigma \exp \left(-\alpha z_i \right)
$$
where $\sigma$ is the baseline length scale and $\alpha$ controls sensitivity to population density. Finally, we define the covariance between $\bv{r}_1$ and $\bv{r}_2$ to be the convolution of two Gaussian kernels with length scales $l_1$ and $l_2$ 
$$
K_s(\bv{r}_1,\bv{r}_2) = \left(\frac{2 l_1 l_2}{l_1^2+l_2^2}\right) \exp \left( - \frac{\Vert \bv{r}_1-\bv{r}_2 \Vert^2}{l_1^2+l_2^2} \right).
$$
The convolution of two Gaussian kernels may be shown to be positive definite \cite{pac06}. In summary, the prior has four parameters, $\tau, \sigma, \alpha, \kappa$ controlling the time-scale ($\tau$) and length-scale ($\sigma$) of fluctuations, the sensitivity ($\alpha$) of length scales to population density, and the overall magnitude  of fluctuations ($\kappa$).  

\subsection{MAP estimation}

In order to obtain MAP estimates of the latent fields for a given linguistic variable we require an expression for the probability of obtaining our observed survey data for that variable, given the values of the latent fields. Since we have a discrete set of birth years we can write the observation times within any interval as the vector $\bv{t}=(t_1, \ldots, t_T)$. Let $F$ be the tensor with components $F_{ijk}=F_{ik}(t_j)$. Also let $y_{ijk}$ be number of speakers (survey respondents) in cell $i$ with birth year $t_j$ using variant $k$, and let $Y$ be the tensor with components $y_{ijk}$. Finally, let $n_{ij}$ be the number of respondents in cell $i$ with birth year $t_j$. The probability of observing $Y$ given $F$ (the likelihood) is then
$$
p(Y|F) = \prod_{i=1}^N \prod_{j=1}^T n_{ij}! \prod_{k=1}^K \frac{\tx{softmax}(\bv{F}_i(t_j))_k^{y_{ijk}}}{y_{ijk}!}.
$$
Writing the prior density of the latent fields as $p(F)$ then the posterior density satisfies $p(F|Y) \propto p(Y|F)p(F)$ and we estimate the latent fields as
$$
\hat{F} = \amax{F}\ p(Y|F) p(F).
$$
This is a challenging optimisation problem due to its high dimensionality. With fifty survey years, four thousand cells and K variants we must infer $2K \times 10^5$ field values, and the prior for each variant is a $2\times 10^5$ dimensional multivariate normal distribution. By exploiting the factorised form of the prior covariance function, and using vectorised evaluation of the posterior and its Jacobian, we can obtain MAP fields in $\approx 1 \tx{min}$ on a modern workstation. 

The results of the above procedure depend on the hyperparameter vector $\bv{\psi} = (\tau, \sigma, \alpha, \kappa)$. We select $\bv{\psi}$ by using cross validation to measure how well the MAP fields describe sample data which was not used for their estimation. We first perform a randomized 80-20 split of the survey data into a training and test set, and calculate the data tensors $Y^{\tx{train}}$ and $Y^{\tx{test}}$. For given $\bv{\psi}$ we compute the MAP state field tensor $\hat{F}^{\tx{train}}$ based on $Y^{\tx{train}}$. The performance of this field estimate as a description of the true empirical fields (those that would be obtained by observing \textit{all} speakers) is measured by the log probability of the test data according to our field estimates $\log p(Y^\tx{test}|\hat{F}^{\tx{train}})$. Up to a field-independent additive constant this is the negative of the categorical cross entropy
$$
\tx{CE}(\bv{\psi}) = - \sum_{i,j,k} Y^{\tx{test}}_{ijk}
\log\left[\tx{softmax}\left(\hat{\bv{F}}^{\tx{train}}_i(t_j)\right)_k\right].
$$
We select the optimal hyperparameter vector which minimises the categorical cross entropy
$$
\hat{\bv{\psi}} = \amin{\bv{\psi} }\ \tx{CE}(\bv{\psi}).
$$
Since each evaluation of the $\tx{CE}$ takes $\approx 1$min, we tackle the optimisation problem using a hyperparameter tuning toolkit (Optuna \cite{aki19}) designed for optimising expensive-to-evaluate objective functions. We restrict the search to the `plausible' hyperparameter region $\tau  \in [20,50](\tx{yrs})$, $\sigma \in [300,500] (\tx{km})$, $\alpha \in [0,0.5]$, $\kappa^2 \in [0.56,1.56]$. Here we set the minimum spatial length scale $\sigma_{\tx{min}}=300$km (just over the largest cell diameter), allowing the population density sensitivity parameter $\alpha$ to shrink the length scale in densely populated regions. We impose the upper limit $\alpha_{\tx{max}}=0.5$ corresponding to a minimum length scale $l_{\tx{min}} \approx 67$km (about half the length of Long Island) in the most densely populated areas.  Without $\sigma_{\tx{min}}$ the preponderance of data in dense regions would yield hyperparameters optimised disproportionately in favour of accuracy in these regions, at the expense of accuracy in sparse regions. Having obtained $\hat{\bv{\psi}}$ for each variable, we use the complete dataset to estimate the (apparent) time evolution of its state field. 

\section{Migration model fitting}
\label{app:migration}

Letting $P_j$ be the  population of $\bv{r}_j$ we assume $w_{ij}$ obeys the following gravity-type model \cite{zip46,sim12}
$$
w_{ij} = c P_i^{\alpha-1} P_j^\alpha (d_0 + d_{ij})^{-\gamma_0 - \gamma_1 \log(d_0+d_{ij})},
$$
where $c>0$, $\alpha \in [0,1]$, $d_0>1$, $\gamma_0>0, \gamma_1 \in \RR$. We ignore within-cell migration since it does not change the state field. Hence we set $w_{ii}=0$. The total migration rate per speaker out of cell $i$ is
$ w_i = \sum_j w_{ij}$, with transition rates between cells depending on their separations and their respective populations. The constant $c$ scales the overall flow rate between all cells. 

 According to this model, the total flow rate of migrants from $\bv{r}_i \ra \bv{r}_j$ is
\begin{equation}
f_{ij} = P_i w_{ij}  =  c (P_i P_j)^\alpha (d_0 + d_{ij})^{-\gamma_0 - \gamma_1 \log(d_0+d_{ij})}.   
\label{eqn:grav}
\end{equation}
Since $f_{ij}=f_{ji}$ then expected cell populations remain constant in this model.  

\begin{figure}
    \centering
    \includegraphics[width=\linewidth]{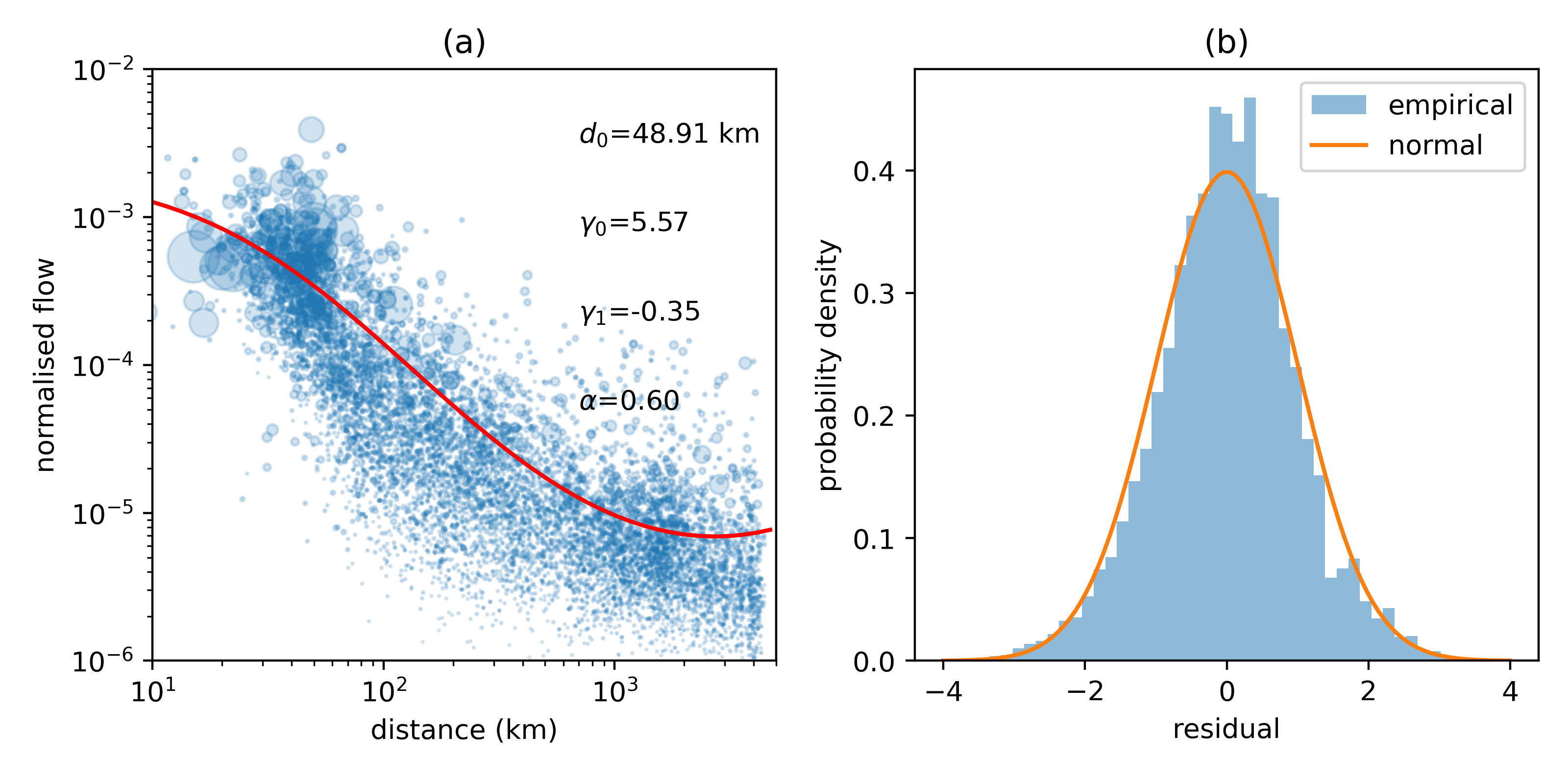}
    \caption{ (a) Scatter plot of normalised 2011 county-county migration flow rates $f_{ij}/(P_iP_j)^\alpha$ between cells with point sizes proportional to absolute flow rates. Red curve shows inferred flow rate model, with parameter values annotated on the plot. (b) Histogram of residuals between the observed and modelled log normalised flows. Orange curve is a normal density fitted to the histogram.  }
    \label{fig:app-mig}
\end{figure}

We estimate parameters using modern county-county migration data and then adjust $c$ to be consistent with historical migration rates \cite{irs11,cen23}. Population estimates for all 3244 US counties are available from the US Census Bureau \cite{cen20}. The US Internal Revenue Service (IRS) provides detailed estimates of the number of individuals per year who have moved residence between each pair of counties based on tax returns. Therefore, we have access to estimates for all flows $f_{ij}$, populations $P_i$ and distances $d_{ij}$. Defining the normalised flow rate $y_{ij} = f_{ij}/(P_iP_j)^\alpha$ the residual between the observed and predicted log normalised flow is
$$
r_{ij} = \log y_{ij} - \log c + \gamma_0 \log(d_0 + d_{ij}) + \gamma_1 \log^2(d_0 + d_{ij}).
$$
We estimate the parameters by minimising the sum of squared residuals using IRS data for 2011 (the earliest available clean dataset), weighted by the flows. Figure \ref{fig:app-mig}a shows the fitted model,  empirical flows and parameter estimates. Figure \ref{fig:app-mig}b shows the distribution of residuals, sampled in proportion to flow size. The near-normality of this distribution justifies our weighted least squares fitting procedure.  In order to fit the constant $c$ to historical migration rates we define the average rate per person
$$
\lambda =  \frac{1}{P} \sum_{i,j}  f_{ij}
$$
where $P$ is total population. During the period [1950,2000] (excluding years 1972–75 and 1977–80 for which the CPS 1-year mobility question was not asked) this rate remained stable \cite{cen24} with mean $\bar{\lambda} = 6.3\%$ (per person per year probability of moving county) and standard deviation $0.4\%$. We select $c$ so that $\lambda=\bar{\lambda}$.

\section{Statistical field model details}

\label{app:SF}

\subsection{Expected state field increment due to variant diffusion}
\label{app:diffusion-derivation}

Here we provide a derivation of the expected state field increment due to variant diffusion, given by
\begin{equation}
\EE_t(\delta \bv{X}_i(t)) = \sum_j \left( w_{ij} + J l_{ij} - (w_i + J) \delta_{ij} \right) \bv{X}_j(t) \dt
\label{eqn:app-dx-diff}    
\end{equation}
where
$$
\bv{X}_i(t) = \frac{1}{P_i} \sum_{s=1}^{P_i} \bv{V}_{is}.
$$
We first consider migration. The total flow rate (in and out) of cell $i$ is $f_i = P_i w_i$. The probability that the speaker in site $k$ in cell $i$ will migrate out in time $\dt$ is then $w_i \dt$. Assuming that this speaker is immediately replaced with an incoming migrant from another cell, then the state of the incoming  speaker will be selected from the probability vector
$$
\bv{p}^{\tx{mig}}_i = \frac{1}{f_i} \sum_j f_{ji} \bv{X}_j(t)  = \frac{1}{w_i} \sum_j w_{ij} \bv{X}_j(t)
$$
where we made use of the fact that $f_{ji}/f_i = f_{ij}/f_i = w_{ij}/w_i$. The conditional expectation of the state of the speaker in site $k$ of cell $i$ at time $t + \dt$, given the state of the system at time $t$, is then
$$
\EE_t(\bv{V}_{is}(t+\dt)) = (1-w_i \dt) \bv{V}_{is}(t) + \bv{p}^{\tx{mig}}_i w_i \dt
$$
where $\EE_t$ denotes expectation conditional on the history of the system up to time $t$. The conditional expectation of $\delta \bv{X}_i(t) = \bv{X}_i(t+\dt) - \bv{X}_i(t)$ due to migration is then
$$
\EE_t(\delta \bv{X}_i(t)) = \sum_j  (w_{ij} - w_i \delta_{ij}) \bv{X}_j(t) \dt.
$$
Now consider copying. The probability that a speaker will copy another speaker's variant during time $\dt$ is $J \dt$. The probability distribution of the copied variant will be
$$
\bv{p}^{\tx{copy}}_i = \sum_j l_{ij} \bv{X}_j(t)
$$
The conditional expectation of $\delta \bv{X}_i(t) = \bv{X}_i(t+\dt) - \bv{X}_i(t)$ due to copying is then
$$
\EE_t(\delta \bv{X}_i(t)) = \sum_j  J(l_{ij} - \delta_{ij}) \bv{X}_j(t) \dt.
$$
Combining the migration and copying increments we obtain equation (\ref{eqn:app-dx-diff}).

\subsection{Construction of lifting matrix}
\label{app:lifting}

The lifting matrix $A$ gives the bias field $\bv{S}$ in terms of the latent field $\bv{\Psi}$ as
\begin{equation}
    \label{eqn:app-Slift}
\bv{S} = A\bv{\Psi}.
\end{equation}
To construct the lifting matrix we first define the Gram matrix \cite{mur22}, $G$, of the centroid vectors of our cells using the radial basis function (Gaussian) kernel with length scale $\eta$
$$
G_{ij} = \exp\left(-\frac{\Vert \bv{r}_i-\bv{r}_j\Vert^2}{2 \eta^2} \right).
$$
Letting $D$ be a diagonal matrix with $D_{ii}=\sum_{j=1}^N G_{ij}$ we define the symmetric degree normalised Gram matrix $\Sigma = D^{-1/2} G D^{-1/2}$ \cite{chu97} and the corresponding Laplacian matrix $L = I - \Sigma$, with eigenvectors $\bv{a}_1, \ldots, \bv{a}_N$, and eigenvalues $\lambda_1, \ldots, \lambda_N$. We define the lifting matrix in terms of these vectors as
\begin{equation}
A = \begin{pmatrix}
| & | & & | \\
\bv{a}_1 & \bv{a}_2 & \ldots &\bv{a}_Q \\
| & | & & |  
\end{pmatrix}.
\label{eqn:lift}    
\end{equation}
Projecting realisations of the Gaussian vector $\bv{V} \sim \N(\bv{0},\Sigma)$, which fluctuates over length scales of order $\eta$, onto the subspace spanned by $\{\bv{a}_i\}_{i=1}^Q$ retains, on average, a fraction
$$
R^2(Q) = \frac{Q- \sum_{k=1}^Q \lambda_k}{N - \sum_{k=1}^N \lambda_k}
$$
of the total variance of its components. Conversely, by choosing $Q$ large enough so that $R^2(Q)$ is close to one, we can represent arbitrary fitness landscapes which fluctuate over length scale $\eta$ in the form (\ref{eqn:app-Slift}).

\subsection{Derivation of stochastic term}
\label{app:stochastic-term}

We will derive the statistical properties of the noise increment in 
\begin{align}
d\bv{X}_i &=  \left(\bv{q}_i \circ \bv{X}_i(t)- \bar{q}_i \bv{X}_i(t) + \sum_j \left( w_{ij} + J l_{ij} - (w_i + J ) \delta_{ij} \right) \bv{X}_j(t)\right)dt + d\bv{\xi}_i(t) \\
&= \bv{b}_i\left(X;\bv{\theta}\right)dt + d \bv{\xi}_i.
\label{eqn:app-dX-full}
\end{align}
under the assumption that variant switching occurs much faster than diffusion and is therefore the dominant contribution to the stochastic term. In this regime $\bar{q}_i \gg \sum_j (w_{ij}+Jl_{ij})$. We also assume that variations in fitness are small compared to average fitness. If this were not the case then the rate at which replicator dynamics took effect would be much faster than diffusion, making diffusion irrelevant to the dynamics. According to this assumption, variant fitness values can be expressed as $q_{ik}=q(1+\epsilon_{ik})$. Letting  $\epsilon \doteq \max_{i,k}|\epsilon_{ik}| \ll 1$ we have
$$
\EE(\delta \bv{V}_{is} \delta \bv{V}_{is}^T|\bv{V}_{is}=\bv{e}_a, \bv{X}_i) = q \dt \sum_{k=1}^K X_{ik}(\bv{e}_k-\bv{e}_a)(\bv{e}_k-\bv{e}_a)^T + O(\epsilon).
$$
Since we also have
$$
\sum_{k=1}^K X_{ik} \bv{e}_k\bv{e}_k^T = \tx{diag}(\bv{X}_i),\quad \sum_{k=1}^K X_{ik} \bv{e}_k = \bv{X}_i, \quad \sum_{k=1}^K X_{ik} = 1
$$
where $\tx{diag}(\bv{X}_i)$ is the diagonal matrix with diagonal entries given by the components of $\bv{X}_i$, and $\EE(\delta \bv{V}_{is})=O(\dt)$ then
$$
\tx{Cov}(\delta \bv{V}_{is}|\bv{V}_{is}=\bv{e}_a,\bv{X}_i) = q \dt \left( \tx{diag}(\bv{X}_i) - \bv{X}_i \bv{e}_a^T - \bv{e}_a \bv{X}_i^T + \bv{e}_a\bv{e}_a^T\right) + O(\dt^2,\epsilon).
$$
Observing that $\PP(\bv{V}_{is}=\bv{e}_a|\bv{X}_i)=X_{ia}$, we have
$$
\EE\left(\tx{Cov}(\delta \bv{V}_{is}|\bv{V}_{is},\bv{X}_i)|\bv{X}_i\right) = 2q \dt \left( \tx{diag}(\bv{X}_i) - \bv{X}_i \bv{X}_i^T\right) + O(\dt,\epsilon). 
$$
Finally, since $\delta \bv{X}_i = P_i^{-1} \sum_{s=1}^{P_i} \delta \bv{V}_{is}$, and according to our assumptions $\tx{Cov}(\delta \bv{\xi}_i) = \tx{Cov}(\delta \bv{X}_i)$, then
$$
\tx{Cov}(d\bv{\xi}_i) \approx \frac{2q}{P_i} \left(\tx{diag}(\bv{X}_i) - \bv{X}_i\bv{X}_i^T\right)dt.
$$
It is important to emphasise that because we are not aiming to infer the typical rate, $q$, of variant switching, the magnitude of the noise term will remain unknown. 

\subsection{Dynamics of modelled variants derived from all-variant model}
\label{app:background}

As stated in the main paper, the survey data to which our model was fitted contained a number of rare variants, which we excluded from the model. We will refer to the $K$ modelled variants as the \textit{main variants} and to the remaining as \textit{background variants}. We show here how our main-variant selection dynamics may be derived from the corresponding all-variant selection dynamics. Since the argument below holds independently at every cell and time we suppress the cell subscript $i$. Let $\hat{X}_k$ be the frequency of variant $k$ in the all-variant case (none excluded) with $\hat{X}_0(t)$ the frequency of variants which we intend to exclude. Let $\hat{\bv{X}} = (\hat X_0,\ldots,\hat X_K)$ denote the all-variant (full-population) frequency vector. Define the total mass of the modelled variants
$$
M(t) = \sum_{k=1}^K \hat X_k(t) = 1-\hat X_0(t).
$$
The relative frequencies of the modelled variants are then given by
$$
X_k(t) = \hat X_k(t)/M(t).
$$
Because respondents reporting an excluded variant are dropped before survey frequencies are computed, $\bv{X}=(X_1,\ldots,X_K)$ is  the paper's state field: a frequency vector conditional on the respondent having used one of the $K$ modelled variants. 

Letting $\hat{\bv{q}}= (\hat{q}_0(\hat{\bv{X}}), \ldots, \hat{q}_K(\hat{\bv{X}}))$ denote the fitness vector in the all variant model, then in the large-population limit we obtain the replicator equation
$$
\frac{d\hat X_k}{dt} = \hat X_k\left(\hat q_k(\hat{\bv{X}}) -
\bar{\hat q}(\hat{\bv{X}})\right), \qquad \bar{\hat q}(\hat{\bv{X}}) =
\sum_{j=0}^K \hat q_j(\hat{\bv{X}}) \hat X_j.
$$
Writing $\ln X_k = \ln \hat X_k - \ln M$, we have
$$
\frac{d\ln \hat X_k}{dt} = \hat q_k(\hat{\bv{X}})-\bar{\hat
q}(\hat{\bv{X}}), \qquad \frac{d \ln M}{dt} = \frac{1}{M}\sum_{j=1}^K
\hat X_j\left(\hat q_j(\hat{\bv{X}})-\bar{\hat q}(\hat{\bv{X}})\right) =
\tilde{\bar q}(\hat{\bv{X}}) - \bar{\hat q}(\hat{\bv{X}}),
$$
where $\tilde{\bar q}(\hat{\bv{X}}) = \sum_{j=1}^K X_j \hat
q_j(\hat{\bv{X}})$ is the mean fitness under the conditional weights
$X_j$. Subtracting, we obtain
\begin{equation}
\frac{dX_k}{dt} = X_k\left(\hat q_k(\hat{\bv{X}}) - \tilde{\bar
q}(\hat{\bv{X}})\right), \qquad k=1,\ldots,K.
\label{eqn:bg_reduction}
\end{equation}
The background fitness $\hat q_0(\hat{\bv{X}})$ has cancelled entirely, regardless of how many excluded variants it aggregates or how their fitnesses vary. This is consistent with the fitness vector's
gauge invariance: shifting $\hat q_k \to \hat q_k + c(\hat{\bv{X}})$ for an arbitrary scalar function $c$ leaves $\hat q_k - \tilde{\bar q}$
unchanged, matching the invariance of the main text's replicator equation under the same transformation.

Equation (\ref{eqn:bg_reduction}) coincides with the paper's replicator equation provided $\hat q_k$ is itself a function of $\bv{X}$ alone. Bias $s_k$ satisfies this trivially. For accommodation,
two readings are possible. If accommodation responds to the modelled-variant frequency actually reported, $q_k(\bv{X})=s_k+\beta
X_k$ as used in the main text, then (\ref{eqn:bg_reduction}) already coincides exactly with the paper's replicator equation for any behaviour of the background. If instead accommodation responds to the true frequency a speaker hears, including background-variant speakers, then $\hat q_k(\hat{\bv{X}}) = s_k + \beta \hat X_k = s_k + \beta M(t)
X_k$, and (\ref{eqn:bg_reduction}) becomes
$$
\frac{dX_k}{dt} = X_k\left[(s_k-\bar s) + \beta M(t)\left(X_k -
\sum_j X_j^2\right)\right], \qquad \bar s = \sum_j X_j s_j,
$$
again the paper's replicator equation, with $\beta$ replaced by an effective, time-varying $\beta_{\rm eff}(t) = \beta M(t)$. The bias term in the conditional selection dynamics is unchanged under either reading. Under the first reading, the accommodation term in the conditional selection dynamics is also unchanged.  Under the second, a single-window fitted $\beta$ is more precisely interpreted as an effective, window-averaged value of $\beta_{\rm eff}(t)=\beta M(t)$; the decade-by-decade estimates of section 4(c) likewise represent effective accommodation factors within each fitting window.

Applying It\^o's lemma to the leading Wright--Fisher diffusion derived above gives, for $M_i>0$,
$$
\tx{Cov}(d\bv{X}_i\mid\hat{\bv{X}}) = \frac{2q}{P_iM_i}\left(\tx{diag}(\bv{X}_i)-\bv{X}_i\bv{X}_i^T\right)dt.
$$
The It\^o drift correction cancels, while the covariance retains the Wright--Fisher form with its amplitude multiplied by $1/M_i$.

\printbibliography

\end{document}